\documentclass[prb,preprint]{revtex4}
\usepackage{amsmath}
\usepackage{graphicx}
\usepackage[latin1]{inputenc}
\usepackage{pstricks}
\usepackage{pst-node}
\usepackage{pst-coil}
\usepackage{pst-grad}
\usepackage{multido}

\begin{document}

\title{Image method in the calculation of the van der Waals force between an atom and a conducting surface}

\author{Reinaldo de Melo e Souza}
\email{reinaldo@if.ufrj.br}
\author{W.J.M. Kort-Kamp}
\email{kortkamp@if.ufrj.br}
\author{C. Sigaud}
\email{sigaud@if.ufrj.br}
\author{C. Farina}
\email{farina@if.ufrj.br}

\affiliation{Universidade Federal do Rio de Janeiro,
Instituto de F\'isica, Rio de Janeiro, RJ 21945-970}


\begin{abstract}

Initially, we make a detailed historical survey of van der Waals forces, collecting the main references on the subject. Then, we review a method recently proposed by Eberlein and Zietal to compute the dispersion van der Waals interaction between a neutral but polarizable atom and a perfectly conducting surface of arbitrary shape. This method has the advantage of relating the quantum problem to a corresponding classical one in electrostatics so that all one needs is to compute an appropriate Green function. We show how the image method of electrostatics can be conveniently used together with the Eberlein and Zietal mehtod (when the problem admits an image solution). We then illustrate this method in a couple of simple but important cases, including the atom-sphere system. Particularly, in our last example, we present an original result, namely, the van der Waals force between an atom and a boss hat made of a grounded conducting material.

\end{abstract}

\maketitle

\section{Historical survey and main purposes}

Intermolecular forces have been studied for approximately three centuries. Since molecules of a real gas condense into liquids and freeze into solids, it is natural to expect that there must exist attractive intermolecular forces, a conclusion that had already been achieved by Newton at the end of the 17th century \cite{Israelachvili-2011}.
%
%
The phenomenon of capillarity - the ability of a liquid to climb the walls of a tube in opposition to external forces like gravity - was studied by the first time by Clairaut who, in 1743, suggested that this phenomenon could be explained if the forces between the molecules of the liquid and those of a tube of glass were different from the intermolecular forces between the molecules of the liquid themselves \cite{Margenau-Kestner-1969}. This same phenomenon was considered later on by Laplace, in 1805, and by Gauss, in 1830. Many other renowned scientists of that time were also involved with the study of determining the force law of intermolecular forces as, for instance, Maxwell and (a little bit later) Boltzmann. Both of them worked in the context of the kinetic theory of gases and based their conclusions on the available data for diffusion coefficients, specific heats and viscosities. Curiously, while Maxwell concluded that the intermolecular forces should be attractive, Boltzmann showed that repulsive forces could explain the available data as well. A more complete list of names involved direct or indirectly to this subject up to the 20th century can be found in the recent edition of the book by Israelachvili \cite{Israelachvili-2011}.

Following a different approach, J.D. van der Waals suggested in his dissertation presented in Leiden in 1873 \cite{VanDerWaals-1873} an equation of state for real gases, given for one mol of gas by $(P + a/V^2)(V - b) = RT$, where $P$, $V$ and $T$ are respectively the pressure, volume and absolute temperature of the gas, $R$ is the universal constant of gases, and $a$ and $b$ two (experimentally) adjustable parameters. Parameter $b$ was introduced by so that the finite volume of the molecules was taken into account: after all, the gas can not be indefinitely compressed to zero volume (as allowed by the equation of state for an ideal gas). On the other hand, the term $a/V^2$ is related to the existence of an attractive intermolecular force, since its presence leads to a smaller pressure. In other words, in a real gas the pressure is smaller than in an ideal gas due to the attractive intermolecular forces. These attractive forces are called, generically, van der Waals forces and since the seminal work of van der Waals much effort has been devoted to understand the nature of such forces. It is worth mentioning that J.D. van der Waals was awarded with the Nobel Prize of Physics in 1910.

In the beginning of the 20th century, M. Reinganum \cite{Reinganum-1903,Reinganum-1912} described the van der Waals forces as a result of the interaction between the permanent electric dipoles of the molecules (he believed that all molecules possessed permanent dipoles). Though this is not true, since there are non-polar molecules, his work can be considered an important step towards the correct interpretation of van der Waals forces. In fact, we must distinguish three types of van der Waals forces: the orientation force, the induction force and the dispersion force, to be described below.

Orientation forces occur between two molecules with permanent electric dipoles. Making a thermal average of the electrostatic interaction between two randomly oriented electric dipoles of moments ${\bf p}_1$ and ${\bf p}_2$, Keesom \cite{Keesom-1915,Keesom-1920} computed the van der Waals interaction energy between two polar molecules in a thermal bath at a given temperature and found
\begin{equation}
U_{or}(r) = -\, \frac{2p_1^2  p_2^2}{3 k_B T (4\pi\epsilon_0)^2 r^6}\; ;\;\;
\mbox{for}\;\; k_B T \gg \frac{p_1 p_2}{4\pi\epsilon_0 r^3}\; ,
\end{equation}
where $p_1 = \vert{\bf p}_1\vert$, $p_2 = \vert{\bf p}_2\vert$, $r$ is the distance between the two molecules, $k_B$ is the Boltzmann constant and $T$ is the absolute temperature. The minus sign in the previous equation means that the interaction is attractive and the subscript {\it or} stems for orientation forces. Though there are as many configurations which give rise to attractive forces as configurations which give rise to repulsive forces, Boltzmann weight ($e^{-{\cal E}/k_B T}$) favours the lower energies which correspond to \lq\lq attractive configurations{\rq\rq}. The presence of $k_B T$ in the denominator is also quite natural, since as $T$ increases indefinitely all configurations (attractive or repulsive ones) become equally available, leading to a vanishing force.

Induction forces are those that occur between a non-polar but polarizable molecule and another one which possess a permanent electric dipole (or even with a higher multipole moment, as for instance an electric quadrupole). Evidences that  non-polar molecules indeed existed, together with the fact that for these substances the van der Waals parameter $a$ was related to the refractive index, led Debye \cite{Debye-1920,Debye-1921} and others to consider this kind of forces. The permanent electric dipole of one molecule  induces an electric dipole in the non-polar but polarizable one, so that we expect in this case also a behavior similar to the previous dipole-dipole interaction. In fact, simple arguments show that the resulting interaction energy for this case will be attractive and proportional to $1/r^6$. If $p_1$ is the magnitude of the dipole moment of the polar molecule then the magnitude of the corresponding electric field at the position of the second (non-polar but polarizable) molecule will be $E_1(2) \sim p_1/r^3$ and, consequently, the magnitude of the induced electric dipole acquired by the second molecule will be $p_2 = \alpha_2 E_1(2)$, where $\alpha_2$ is the static polarizability of the second molecule. Hence, apart from a simple numerical factor, we get for the interaction energy $U_{ind} \sim - p_2E_1(2) \sim -\alpha_2 p_1^2/r^6$. Though the spatial dependence of  $U_{ind}(r)$ is the same as that for  $U_{or}(r)$, the induction force does not disappear for high temperatures, since the orientation of the dipoles 1 and 2
are not independent as in the orientation force. Indeed, in a first approximation, the induced dipole 2
is parallel to the field generated by the dipole 1 at the position occupied by the dipole 2, which also explains the attractive character of the induction van der Waals interaction between a polar molecule and a non-polar one.

The two types of van der Waals forces just described can not be used to explain the attraction between two non-polar molecules or two atoms, like those of noble gases (a clear evidence that these forces indeed exist lies in the fact that noble gases condense). In other words, the term $a/V^2$ in van der Waals equation of state will still be necessary to describe noble gases more accurately than the ideal gas equation of state does. However, the correct explanation for the forces between two atoms or two non-polar molecules, called dispersion van der Waals forces, had to wait for the advent of Quantum Mechanics. Due to quantum fluctuations, the charge and current distributions  in an atom (or molecule) fluctuate  and consequently  we can think that instantaneous dipoles (or higher multipoles) exist and give rise to an electromagnetic interaction. These quantum fluctuations are ultimately related to the Heisenberg uncertainty principle, one of the most important pillars of Quantum Mechanics. In 1927, Wang \cite {Wang-1927} solved the Schrödinger equation for two Hydrogen atoms separated by a distance $r$ much greater than the Bohr radius $a_0$ but considering the contribution of the instantaneous electric dipole interaction potential between the two atoms. After using a complicated perturbation method developed by Epstein \cite{Epstein-1926,Epstein-1927}, he found
\begin{equation}
U_{Wang}(r) \approx -\, 8.7\,\frac{e^2 a_0^2}{(4\pi\epsilon_0)^2 r^6}\, .
\end{equation}
Three years later, Eisenschitz and London \cite{Eisenschitz-London-1930} and London \cite{London-1930} considered the same problem in much more detail and used a much simpler perturbative method, refining the numerical factor in the previous result and relating the interaction potential for the dispersion van der Waals forces directly to the atomic polarizability of the Hydrogen atom. By the way, since the dynamical polarizability $\alpha(\omega)$ is closely related to the permittivity $\epsilon(\omega)$, these forces are called dispersion van der Waals forces (name coined by London in the latter article \cite{London-1930}). The expression for the dispersion interaction energy between two polarizable atoms can be written in the form \cite{London-1930}
\begin{equation}\label{Disp}
U_{disp}(r) = -\frac{3}{4}\frac{\hbar\omega_0\alpha_0^2}{(4\pi\epsilon_0)^2 r^6}\, ,
\end{equation}
where $\omega_0$ is the dominant transition frequency for the interaction and $\alpha_0$ is the corresponding static polarizability of the atoms. Although the $(-1/r^6)$ power law already appears in the first paper \cite{Eisenschitz-London-1930}, the second one, written solely by London  \cite{London-1930}, is by far more cited than Eisenschitz and London's paper \cite{Eisenschitz-London-1930}, so that dispersion van der Waals forces are usually called London forces (an ordinary quantum mechanical calculation of London's result can be found in many textbooks, as for instance those written by Cohen-Tannoudji, Diu and Laloë \cite{CohenEtAl-1973} and by Bransden and Joachain \cite{Bransden-Joachain-2000}).

For more than one decade after London's result had been established, people believed that the final explanation for dispersion forces was that given by London. However, experiments with colloidal suspensions (for a comprehensive  discussion on colloids we suggest the nice book by J. Berg \cite{Berg-Book-2010}), made in the Phillips laboratories in the first half of the 1940's by Verwey  and Overbeek, showed that if London's result were used there would be discrepancies between theoretical predictions and experimental data (see for instance the paper by Verwey \cite{Verwey-1947} and the book by Verwey and Overbeek \cite{Verwey-Overbeek-1948}). They noticed that in order to retrieve agreement between experimental data and theory the dispersion interaction energy between two atoms (or two non-polar but polarizable molecules) should fall for large distances more rapidly than $1/r^6$. Further, Overbeek conjectured that such a change in the force law was due to retardation effects of the electromagnetic interaction, since the velocity of light is finite. Retardation effects become important as the time elapsed by light to propagate from one atom to the other is of the order of characteristic times of the atoms, namely, $1/\omega_{mn}$, where $\omega_{mn}$ are the allowed transition frequencies of the atoms. Assuming there is a dominant transition frequency, say $\omega_0$, retardation effects cease to be negligible for $r/c \ge 1/\omega_0$ (in terms of wavelengths, this condition is written as $r\ge \lambda_0$, where $\lambda_0$ is the wavelength of the dominant transition). Generally speaking, we can then distinguish two regimes for dispersion interactions, namely: the non-retarded or short distance regime and the (asymptotically) retarded or large distance regime. The latter is valid for $r\gg \lambda_0$ while the former is valid for $a_0\ll r\ll \lambda_0$, with $a_0$ being the Bohr radius (the condition $a_0\ll r$ is to avoid the overlapping of the electronic clouds of the two interacting atoms). The influence of retardation effects on the London-van der Waals forces was first reported by Casimir and Polder  in 1946 in a very short paper \cite{Casimir-Polder-1946}. Two years later,  Casimir and Polder published a large paper containing all the details of a fourth order perturbative calculation in Quantum Electrodynamics that led them to their previous results \cite{Casimir-Polder-1948}. They showed that in the asymptotically retarded regime $(r\gg \lambda_0)$, the dispersion interaction energy between two polarizable atoms is given by
\begin{equation}
U_{ret}(r) = -\frac{23\hbar c}{4\pi}\frac{\alpha_1\alpha_2}{(4\pi\epsilon_0)^2 r^7}\, ,
\end{equation}
where $\alpha_1$ and $\alpha_2$ are the static polarizabilities of atoms 1 and 2, respectively. As we can see from the previous equation, in the retarded regime the power law of the interaction energy changes by a factor one, from $1/r^6$ to $1/r^7$. In this same paper, Casimir and Polder also showed that the dispersion interaction energy between a polarizable atom and a perfectly conducting plane, in the (asymptotically) retarded regime is proportional to $1/z^4$, where $z$ is the distance from the atom to the conducting plane, instead of proportional to $1/z^3$, a result valid in the non-retarded regime, as showen by the first time by Lennard-Jones in 1932 \cite{Lennard-Jones-1932}. We can understand, qualitatively, why retardation effects weaken the interaction between two atoms as follows. When the time taken for the electric field created by the fluctuating dipole of atom 1 to reach atom 2 and return to atom 1 is of the order of the period of the fluctuating dipole itself, the instantaneous dipole of atom 1 will have changed substantially from its original value so that the mutual configuration of both atoms is less correlated and less favourably disposed for an attractive interaction \cite{Israelachvili-2011}. It is worth mentioning that the orientation and induction van der Waals interactions remain non-retarded at all separations, only the dispersion van der Waals interaction is influenced by the retardation effects of the electromagnetic interactions \cite{Israelachvili-2011}. The first time a transition from the retarded regime to the non-retarded one was observed  occurred only in 1968 in the experiment made by Tabor and Winterton \cite{Tabor-Winterton-Nature-1968,Tabor-Winterton-1968}.

In contrast to what happens to the Coulomb interaction among many point charges, which obeys the so called Superposition Principle,
van der Waals interactions, in general, are not pairwise additive, as first noticed by Axilrod and Teller \cite{Axilrod-Teller-1943}. By non-additivity of van der Waals interactions we mean that the interaction between two atoms is affected by the presence of a third one. This fact must be taken into account in the computation of the van der Waals force between an atom and a macroscopic body (or between two macroscopic bodies), since a pairwise integration using the London or Casimir and Polder forces is not rigorously valid anymore, except for rarefied bodies. The non-additivity of van der Waals forces are ultimately related to multiscattering processes. For instance, in the case of three atoms, the field emanated by the first atom can reach the second one directly or after being scattered by the third atom. Non-additivity effects on the energy of a system may be positive as well as negative and are usually small (approximately less than $20\%$), but they can be very important, as for instance, in the way the atoms of rare gases are arranged in solids \cite{Israelachvili-2011}. More details about the the non-additivity of dispersion van der Waals interaction can be found in Ref(s) \cite{Margenau-Kestner-1969,Langbein-1974,Milonni-1994} (see also Ref. \cite{Farina-Santos-Tort-1999} for a simple way of understanding this feature of dispersion van der Waals forces). Detailed calculations of the dispersion van der Waals interaction between two polarizable atoms at any separation can be found in the pedagogical paper by Holstein \cite{Holstein-2001} and in some textbooks like those of Power \cite{Power-1964}, Craig and Thirunamachandran \cite{Craig-Thiru-1998}, and Salam \cite{Salam-2010}, where a careful analysis of the particular cases of the retarded and non-retarded regimes are presented. For an introductory discussion on the three types of van der Waals interactions and also the dispersion force between a polarizable atom and a conducting sphere see Taddei {\it et al} \cite{Taddei-Mendes-Farina-2010}.

Though we are not going to discuss any experiment on the measurement of dispersion forces involving atoms and macroscopic surfaces, it is worth mentioning a few of them. One of the first experiments involving a beam of atoms scattering by a cylindrical surface was made in 1969 by Raskin and Kusch \cite{Raskin-Kusch-1969}. In 1993, a remarkable experiment was done by Sukenik and collaborators \cite{SukenikEtAl-1993}, in which for the first time the change in the power law between retarded and non-retarded regimes were observed directly with atoms. In 1996, Landragin and collaborators measured the van der Waals force in an atomic mirror based on evanescent waves \cite{LandraginEtAl-1996}. A few years later, quantum reflection was used to measure dispersion forces by Shimizu \cite{Shimizu-2001}. A short but valuable description of these experiments can be found in the nice paper by Dalibard \cite{Dalibard-2002}.

Dispersion forces appear not only in different areas of Physics, as in atomic and molecular physics, condensed matter physics and quantum field theory, but also in Engineering, Chemistry and Biology \cite{Parsegian-Book}. In Quantum Field Theory, it is closely connected to the Casimir effect
(for a short history of this effect and its origin in experiments on colloidal chemistry see the introductory papers \cite{Elizalde-Romeo-1991,Farina-BJP-2006} and the books \cite{Milonni-1994,Proceedings-Leipzig-1998,Milton-Book-2004,Mostepanenko-Book-2009} and references therein). There are even more bizarre situations where dispersion forces play an important role, like in the adhesion of geckos to the ceiling of our houses \cite{AutumnEtAl-Gecko-2002,LeeEtAl-Nature-2007} or in the formation of a thin liquid layer  on  ice that has been identified as an important element in the generation of electric potentials in thunderstorms \cite{Lamoreaux-PhysicsToday-2007}. A vast list of references on this subject can be found in the Resource Letter published in this journal by Milton \cite{Milton-AJP} and a very complete list of references on dispersion forces can be found in the paper by Buhmann and Welsch \cite{Buhmann-Welsch-2007}. Concerning the whole history of intermolecular forces we recommend the excellent book by Rowlinson \cite{Rowlinson-Book-2002}, where the reader can find a huge bibliography containing the relevant works written in the last three centuries

Though the correct description of dispersion forces between two polarizable atoms separated by an arbitrary distance demands, somehow, the quantization of the electromagnetic field, which makes the general calculation very hard, the particular case of non-retarded dispersion forces (or simply dispersion van der Waals forces) can be computed with ordinary quantum mechanics avoiding completely the quantization of the electromagnetic field. In this article, we shall be concerned only with non-retarded dispersion forces. Particularly, we shall focus our attention to the van der Waals interaction energy between a polarizable atom and a perfectly conducting surface. Our purposes here are the following: {\it (i)} to popularize a simple but powerful method  proposed in 2007 by Eberlein and Zietal \cite{Eberlein-Zietal-2007}, and used afterward by them and other authors
 \cite{Eberlein-Zietal-2011,Contreras-Eberlein-2009,ReinaldoEtAl-2011,ReinaldoEtAll-Proceeding-2012}, which is extremely well suited for this kind of calculation; {\it (ii)} to show  that the usual image method of electrostatics can be very useful in applying Eberlein and Zietal method (this will become clear in the explicit solutions of a couple of examples) and {\it (iii)} to present an original result, namely, the calculation of the dispersion van der Waals force between an atom and a conducting \lq\lq boss hat{\rq\rq} (a conducting hemisphere attached to an infinite conducting plane).  Our aim is to make our presentation in a level easily understandable to an undergraduate
student with a good background in elementary electromagnetism and quantum mechanics.

This paper is organized as follows. In the Section 2 we review the  Eberlein and Zietal method showing how we can
combine it with the image method. In Section 3 we illustrate the method by solving explicitly a couple of introductory problems, namely, the simple atom-plane system and the less obvius atom-sphere system. In Section 4 we treat the non-trivial case of an atom in the presence of a boss hat
which, as far as the author's knowledge goes, had  not appeared in the literature before. Section 5 is left for conclusions and final remarks.

\section{Eberlein-Zietal Method}

In this section we review Eberlein and Zietal method \cite{Eberlein-Zietal-2007}. As we shall see, this method has the advantage of relating the quantum problem to a corresponding classical one in electrostatics so that all one needs is to compute an appropriate solution of Laplace equation.

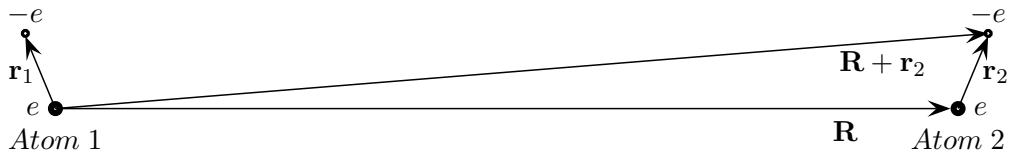
\begin{figure}[!h]
\begin{center}
\newpsobject{showgrid}{psgrid}{subgriddiv=1,griddots=10,gridlabels=6pt}
\begin{pspicture}(-7,0)(7,1.5)
\psset{arrowsize=0.18 2}
\psset{unit=1}

\pscircle[linecolor=black,linewidth=0.8mm](-6,0){.1}
\rput(-6.3,0){$e$}
\rput(-6.45,0.45){${\bf r}_1$}
\psline[linecolor=black,linewidth=0.2mm]{->}(-6,0)(-6.4,0.96)
\pscircle[linecolor=black,linewidth=0.8mm](-6.4,1){.06}
\rput(-6.4,1.25){$-e$}
\rput(-6.0,-0.4){$Atom \; 1$}

\pscircle[linecolor=black,linewidth=0.8mm](6,0){.1}
\rput(6.3,0){$e$}
\rput(6.5,0.45){${\bf r}_2$}
\psline[linecolor=black,linewidth=0.2mm]{->}(6,0)(6.4,0.96)
\pscircle[linecolor=black,linewidth=0.8mm](6.4,1){.06}
\rput(6.4,1.25){$-e$}
\rput(6.0,-0.4){$Atom \; 2$}

\psline[linecolor=black,linewidth=0.2mm]{->}(-6,0)(5.9,0)
\rput(4.5,-0.3){${\bf R}$}

\psline[linecolor=black,linewidth=0.2mm]{->}(-6,0)(6.4,1)
\rput(5,0.6){${\bf R} + {\bf r}_2$}

\end{pspicture}
\end{center}
\caption{Two Hydrogen atoms whose nucleus are separated by a distance $R$ much greater than their sizes. For convenience, the figure is not in scale. The relative position vectors ${\bf R} - {\bf r}_1$ and ${\bf R} + {\bf r}_2 -{\bf r}_1$ are not shown in the figure to avoid overloading it.}
\label{2Atoms}
\end{figure}

However, with the purpose of shedding some light in the method to be explained, we start for convenience by making a few comments on how to compute the van der Waals force between two Hydrogen atoms. FIG. \ref{2Atoms} shows two Hydrogen atoms separated by a distance $R = \vert{\bf R}\vert$, as well as other relative position vectors relevant to the problem (we have not included in this figure the other relative position vectors to avoid overloading it). In order to obtain the interaction hamiltonian, to be used in the perturbative quantum mechanical calculation of the interaction energy between the atoms, we write the electrostatic Coulomb interaction $U_{Coul}$ among all the charges of the whole system. Then,  we subtract from $U_{Coul}$ the coulomb interaction between the electron and the proton of each atom and, finally, we make a Taylor expansion assuming that $r_1, r_2\ll R$, where $r_1 = \vert{\bf r}_1\vert$ and $r_2 = \vert{\bf r}_2\vert$.

The above mentioned Coulomb interaction energy is readly given by
\begin{equation}
U_{Coul} = -\frac{e^2}{4\pi \epsilon_0}\left\{ \frac{1}{r_1} + \frac{1}{r_2}\right\}
+
\frac{e^2}{4\pi \epsilon_0}\left\{\frac{1}{R} - \frac{1}{\vert{\bf R} + {\bf r}_2\vert}
- \frac{1}{\vert{\bf R} - {\bf r}_1\vert}  +  \frac{1}{\vert{\bf R} + {\bf r}_2 -{\bf r}_1\vert}\right\}\, .
\end{equation}
The interaction hamiltonian $H_{int}$ is then easily identified as the sum of the last four terms on the right hand side of the previous equation. It is straightforward to show that, after a Taylor expansion, the dominant contribution for $H_{int}$ can be written as
\begin{equation}
H_{int} = \frac{1}{4\pi\epsilon_0}\left\{
\frac{{\bf p}_1\cdot{\bf p}_2 - 3({\bf p}_1\cdot\hat{\bf R})({\bf p}_2\cdot\hat{\bf R})}{R^3}\right\}\, ,
\end{equation}
where ${\bf p}_1 = e{\bf r}_1$ and ${\bf p}_2 = e{\bf r}_2$ are the quantum mechanical electric dipole operators, and $\hat{\bf R} = {\bf R}/R$. As expected, the dominant contribution is the dipole-dipole interaction. Assuming the two atoms are on their fundamental states, it is easy to check that a first order perturbative calculation yields a vanishing result, due to spherical symmetry of the fundamental wavefunctions. Performing a second order perturbative calculation one obtains \cite{CohenEtAl-1973,Bransden-Joachain-2000} an attractive interaction energy proportional to $1/R^6$,  as discussed in the introduction (see equation (\ref{Disp})).

The lesson to be learned here is the fact that in the non-retarded regime, the interaction hamiltonian is given by the Coulomb energy of the system, after appropriate subtractions. However, for more complicated situations involving continuous distributions of charges, as will be the case of our interest, instead of computing $U_{Coul}$ by a pairwise summation of the coulomb interaction between all pairs of charges, we may  express the electrostatic energy in terms of the electrostatic potential $\Phi({\bf r})$, namely,
\begin{equation}\label{UCoulomb}
U_{Coul} = \frac{1}{2}\sum_{i\atop{i\ne j}}\sum_j\frac{q_i q_j}{4\pi\epsilon_0\, \vert{\bf r}_i - {\bf r}_j\vert}
\;\;\;\Longrightarrow\;\;\;
U_{Coul} = \frac{1}{2}\int \rho({\bf r}) \Phi({\bf r})\, d^3{\bf r}\, .
\end{equation}

Let us now come back to our problem, namely, that of a polarizable atom close to a perfectly conducting surface of arbitrary shape. This leads us naturally to the corresponding classical problem of a dipole near a conducting surface. The presence of the dipole induces a surface distribution of charges and we need to calculate the electrostatic energy of the system in order to  determine the interaction hamiltonian to be used in the subsequent perturbative (quantum mechanical) calculation. As it will become evident, this task will be conveniently made with the aid of equation (\ref{UCoulomb}).

The electrostatic potential $\Phi({\bf r})$  satisfies the Poisson equation
\begin{equation}\label{poisson}
\nabla^2\Phi(\mathbf{r}) = -\frac{\rho(\mathbf{r})}{\varepsilon_0} \, ,
\end{equation}
where $\rho(\mathbf{r})$ is the charge density, submitted to the appropriate  boundary condition on the surface $S$.
 Assuming, initially, that we have a grounded surface, the BC imposed on the electrostatic potential is given by
\begin{equation}\label{ccat}
\Phi(\mathbf{r})\Bigg|_{\mathbf{r}\in\mathcal{S}}=0 \, .
\end{equation}
The electrostatic energy of the configuration is then given by equation (\ref{UCoulomb}). Solutions of equation (\ref{poisson}) can be obtained
by using the Green function method \cite{Byron1992}, where the Green function $G({\bf r},{\bf r}^{\,\prime})$ satisfies, by definition, the following equation
\begin{equation}\label{green}
\nabla^2G(\mathbf{r},\mathbf{r}') = -\delta(\mathbf{r}-\mathbf{r}') \, .
\end{equation}
Therefore, a general solution of equation (\ref{poisson}) can be written as
\begin{equation}\label{phig}
\Phi(\mathbf{r}) = \frac{1}{\varepsilon_0}\int G(\mathbf{r},\mathbf{r}') \rho (\mathbf{r}') d^3\mathbf{r}' \, .
\end{equation}
In order that the electrostatic potential obeys the BC written in (\ref{ccat}) it suffices to impose the same  BC to the
Green function, namely,
\begin{equation}\label{ccgat}
G(\mathbf{r},\mathbf{r}')\Bigg|_{\mathbf{r}\in\mathcal{S}}=0 \, .
\end{equation}
In terms of the Green function $G({\bf r},{\bf r}^{\,\prime})$ the electrostatic energy given by (\ref{UCoulomb}) takes the form
\begin{eqnarray}\label{energyg}
U_{Coul} &=& \frac{1}{2}\int \!\! d^3\mathbf{r}\; \rho(\mathbf{r})\Phi(\mathbf{r}) \cr
&=&
\frac{1}{2\varepsilon_0}\int\!\! d^3\mathbf{r}\, d^3\mathbf{r}'\;\rho(\mathbf{r}) G(\mathbf{r},\mathbf{r}')\rho(\mathbf{r}')  \, .
\end{eqnarray}
One solution of equation (\ref{green}) is readily obtained, since this equation is nothing but the Poisson
equation for a point charge at position $\mathbf{r}'$, except for a constant multiplicative factor. Hence, a particular solution of (\ref{green}) is given by
\begin{equation}
G_p({\bf r},{\bf r}^{\,\prime}) = \frac{1}{4\pi|\mathbf{r}-\mathbf{r}'|}\, .
\end{equation}
However, this solution does not obey the correct BC given by (\ref{ccgat}). In order to adjust the BC, we add to this particular solution a solution of the homogeneous equation and write
\begin{equation}\label{G}
G(\mathbf{r},\mathbf{r}')=\frac{1}{4\pi|\mathbf{r}-\mathbf{r}'|} + G_H(\mathbf{r},\mathbf{r}') \, ,
\end{equation}
where $G_H({\bf r},{\bf r}^{\,\prime})$  satisfies Laplace equation,
\begin{equation}\label{laplace}
\nabla^2G_H(\mathbf{r},\mathbf{r}')=0 \, .
\end{equation}
From (\ref{ccgat}) and (\ref{G}), we immediately determine the BC satisfied by $G_H({\bf r},{\bf r}^{\,\prime})$, which reads
\begin{equation}\label{cch}
\left[\frac{1}{4\pi|\mathbf{r}-\mathbf{r}'|} + G_H(\mathbf{r},\mathbf{r}')\right]_{\mathbf{r}\in S}=0 \, .
\end{equation}
Recall that  all the information about the geometry of the problem is contained in $G_H(\mathbf{r},\mathbf{r}')$.
Let us now consider the charge density to be used in our problem. Regarding the atom as an electric dipole with the positive point charge
at position ${\bf r_0}$ and the negative one at position ${\bf r_0} + {\bf h}$ (in the appropriate moment we will take the limit
 ${\bf h}\rightarrow {\bf 0}$), and recalling that $\Phi({\bf r})$ vanishes on the surface, we may write
\begin{equation}\label{denscargaeb}
\rho (\mathbf{r}) =
 q \Bigl[\delta(\mathbf{r} - \mathbf{r}_0) - \delta\Bigl(\mathbf{r} - (\mathbf{r}_0 + \mathbf{h})\Bigr)\Bigr] \, ,
\end{equation}
%
%
We now substitute (\ref{denscargaeb}) and (\ref{G}) into (\ref{energyg}) and after that we take the limit
$\mathbf{h}\rightarrow 0$, with $q\mathbf{h} \rightarrow \mathbf{d}$ (in a moment this will be interpreted as the atomic dipole operator), we obtain
\begin{eqnarray}\label{taylor1}
U_{Coul} = \lim\limits_{\mathbf{h}\rightarrow 0\atop{ q\mathbf{h}=\mathbf{d}} }
 &\bigg\{ &
 \frac{1}{8\pi\varepsilon_0} \frac{q^2}{|(\mathbf{r}_0+\mathbf{h}) - (\mathbf{r}_0+\mathbf{h})|} \;+\; \frac{1}{8\pi\varepsilon_0}\frac{q^2}{|\mathbf{r}_0-\mathbf{r}_0|} \cr\cr
&+& \frac{1}{8\pi\varepsilon_0}\frac{q^2}{|\mathbf{r}_0+\mathbf{h}-\mathbf{r}_0|} \;+\; \frac{1}{8\pi\varepsilon_0}\frac{q^2}{|\mathbf{r}_0-(\mathbf{r}_0+\mathbf{h})|} \cr\cr
&+&
\frac{q^2}{2\varepsilon_0} [G_H(\mathbf{r}_0+\mathbf{h},\mathbf{r}_0+\mathbf{h}) \;-\; G_H(\mathbf{r}_0+\mathbf{h},\mathbf{r}_0)]\cr\cr
&-&
\frac{q^2}{2\varepsilon_0} [G_H(\mathbf{r}_0,\mathbf{r}_0+\mathbf{h})-G_H(\mathbf{r}_0,\mathbf{r}_0)] \bigg\} \, .
\end{eqnarray}
Although the previous expression for $U_{Coul}$ contains eight terms, only the last four terms are of interest, since only these terms
contain information about the interaction between the dipole and the surface. Indeed, the first two terms account for the divergent self-interaction of the point charges at $\mathbf{r}_0$ and $\mathbf{r}_0 + \mathbf{h}$ and the next two terms  stand for
the divergent self-interaction of the dipole. The remaining terms can be put into a more useful form.
Making a Taylor expansion of $G_H(\mathbf{r}_0 + \mathbf{h},\mathbf{r}_0+\mathbf{h})$ in powers of ${\bf h}$, it follows that
\begin{eqnarray}\label{Taylor1}
	G_H(\mathbf{r}_0+\mathbf{h},\mathbf{r}_0+\mathbf{h})
 &-&
 G_H(\mathbf{r}_0+\mathbf{h},\mathbf{r}_0)\cr
 &=& G_H(\mathbf{r}_0+\mathbf{h},\mathbf{r}_0) +
	+ \mathbf{h}\cdot\nabla'G_H(\mathbf{r}_0+\mathbf{h},\mathbf{r}')\big|_{\mathbf{r}'=\mathbf{r}_0}
\!\!\!- \;G_H(\mathbf{r}_0+\mathbf{h},\mathbf{r}_0) \nonumber \\
&=&\mathbf{h}\cdot\nabla'G_H(\mathbf{r}_0+\mathbf{h},\mathbf{r}')\big|_{\mathbf{r}'=\mathbf{r}_0} \, .
\end{eqnarray}
Analogously, a Taylor expansion of $G_H(\mathbf{r}_0,\mathbf{r}_0+\mathbf{h})$ yields
\begin{equation}\label{Taylor2}
	G_H(\mathbf{r}_0,\mathbf{r}_0+\mathbf{h}) - G_H(\mathbf{r}_0,\mathbf{r}_0) =
 {\bf h}\cdot\nabla G_H(\mathbf{r}_0,\mathbf{r}')\bigg|_{\mathbf{r}'=\mathbf{r}_0}\, .
\end{equation}
Hence, subtracting the irrelevant self-interaction terms mentioned before and using equations (\ref{Taylor1}) and (\ref{Taylor2}), the relevant interaction hamiltonian operator to be used in the perturbative quantum mechanical calculation of the van der Waals force between the atom and the (grounded) conducting surface is given by
\begin{eqnarray}
H_{int}
 &=&
 \lim_{\mathbf{h}\rightarrow 0\atop{q\mathbf{h}=\mathbf{d}} }\frac{q}{2\varepsilon_0}(\mathbf{d}\cdot\nabla')
 \Big[G_H(\mathbf{r}_0+\mathbf{h},\mathbf{r}')-G_H(\mathbf{r}_0,\mathbf{r}')\Big]_{\mathbf{r}'=\mathbf{r}_0} \nonumber \\
\label{mantay}	
&=& \frac{1}{2\varepsilon_0}(\mathbf{d}\cdot\nabla')(\mathbf{d}\cdot\nabla)G_H(\mathbf{r},\mathbf{r}')\bigg|_{\mathbf{r}=\mathbf{r}'=\mathbf{r}_0} \, . \label{taylor2}
\end{eqnarray}
It is worth emphasizing that the atomic dipole moment that appears in this expression is a quantum operator. In first order of perturbation theory the desired non-retarded interaction energy between the atom and the conducting surface, denoted by $U_{NR}$, is just the quantum
expectation value of the above expression, namely,
\begin{equation}
 U_{NR}({\bf r}_0) = \langle H_{int}\rangle = \frac{1}{2\varepsilon_0}\sum\limits_{m,n=1}^{3}\langle d_md_n\rangle\nabla_m\nabla_n' G_H(\mathbf{r},\mathbf{r}')\bigg|_{\mathbf{r}=\mathbf{r}'=\mathbf{r}_0} \, , \label{enpert}
\end{equation}
where we used the fact that the only operators in the above expression are $d_m$ and $d_n$ ($G_H({\bf r},{\bf r}^{\,\prime})$ is a c-number).
 For the sake of simplicity we shall always work with an orthonormal basis, for which we can write
\begin{equation}
\langle d_md_n\rangle=\delta_{mn}\langle d_m^2\rangle \, .
\end{equation}
Combining the two last equations we finally obtain
\begin{equation}\label{eberlein}
U_{NR}({\bf r}_0) = \frac{1}{2\varepsilon_0}\sum\limits_{m=1}^3\langle d_m^2\rangle\partial_m\partial_m^{\,\prime} G_H(\mathbf{r},\mathbf{r}')\big|_{\mathbf{r}=\mathbf{r}'=\mathbf{r}_0} \, ,
\end{equation}
which is precisely the expression obtained by Eberlein and Zietal \cite{Eberlein-Zietal-2007}.
This method has the advantage of relating the quantum problem to a corresponding classical one in electrostatics. Its remarkable simplicity consists in the fact that to obtain the non-retarded van der Waals interaction energy of an atom near \emph{any}
conducting surface one must  find only the homogeneous solution of Laplace equation, $G_H$, corresponding to that geometry. In other works, one must solve the classical problem defined by equations (\ref{laplace}) and (\ref{cch}).
These equations are, except for constants, precisely those that yield the electrostatic potential of
the image charges for the problem of a charge at position $\mathbf{r}'$ in the presence of the surface $\mathcal{S}$,
(if the problem admits a treatment by the image method).

Indeed, denoting by $\phi(\mathbf{r})$ the electrostatic potential of that configuration, we may
decompose it into the sum of the potential of a charge plus the potential of the image charges, denoted by $\phi_i(\mathbf{r})$.
Since the potential of the single charge  satisfies Poisson equation of the problem,
$\phi_i(\mathbf{r})$ obeys Laplace equation. Together with the boundary condition,
the equations satisfied by $\phi_i(\mathbf{r})$ are the following
\begin{eqnarray}
	\label{imagem}
	\nabla^2 \phi_i(\mathbf{r})&=&0 \\
	\label{ghphi}\left[\frac{q}{4\pi\varepsilon_0|\mathbf{r}-\mathbf{r}'|} + \phi_i(\mathbf{r})\right]_{\mathcal{S}}&=&0 \, .
\end{eqnarray}
From a direct comparison between equations (\ref{imagem}) and (\ref{ghphi}) and equations (\ref{laplace}) and (\ref{ccgat}), it is
straightforward to make the identification
\begin{equation}\label{imgh}
G_H(\mathbf{r},\mathbf{r}')=\frac{\varepsilon_0\phi_i(\mathbf{r})}{q} \, .
\end{equation}
Note that the dependence of the rhs of previous equation on $\mathbf{r}'$ is implicit, since the image charge depends on the position of the physical charge - recall that the physical charge is located at $\mathbf{r}'$.

In conclusion, the image method is a very useful tool in order to find the homogeneous solution $G_H$ which, in turn, is the
only function needed to perform Eberlein-Zietal calculation and obtain the quantum non-retarded dispersive interaction
between an atom and a conducting surface $\mathcal{S}$ of an arbitrary shape. In the next sections we will apply this procedure for
different geometries.

\section{Introductory examples}

In this section, we apply Eberlein and Zietal's method just discussed in introductory examples, namely, for an atom close to a grounded conducting plane and an atom close to a grounded conducting sphere, where the image method can be employed with no difficulty. In the latter case, it is  possible to solve for a non-grounded isolated sphere as well. However, in this case, appropriate modifications of the method outlined in the previous section are needed, since  $G_H({\bf r},{\bf r}^{\,\prime})$ no longer satisfies the boundary condition (\ref{ccgat}).

\subsection{Atom close to a grounded conducting plane}

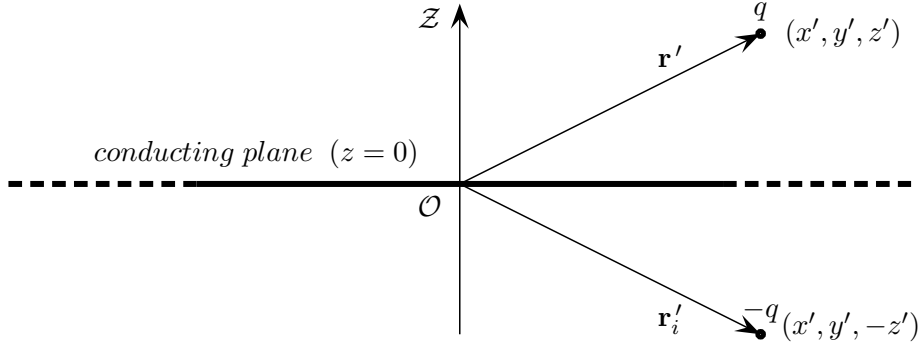
\begin{figure}[!h]
\begin{center}
\newpsobject{showgrid}{psgrid}{subgriddiv=1,griddots=10,gridlabels=6pt}
\begin{pspicture}(-4.5,-1.3)(4.5,2.5)
\psset{arrowsize=0.2 2}
\psset{unit=1}

\psline[linecolor=black,linewidth=0.8mm,linestyle=dashed]{-}(-6,0)(-3.5,0)
\psline[linecolor=black,linewidth=0.8mm]{-}(-3.5,0)(3.5 ,0)
\psline[linecolor=black,linewidth=0.8mm,linestyle=dashed]{-}(3.5,0)(6,0)
\rput(-2.7,0.4){$conducting \; plane\;\;(z=0)$}

\psline[linecolor=black,linewidth=0.2mm]{->}(0,-2.0)(0 ,2.4)
\rput(-0.4,2.2){{${\cal Z}$}}
\rput(-0.4,-0.3){{${\cal O}$}}

\pscircle[linecolor=black,linewidth=0.8mm](4 , 2){.07}
\rput(4,2.3){{$q$}}
\psline[linecolor=black,linewidth=0.2mm]{->}(0,0)(4 , 2)
\rput(5.1,2){{($x', y', z'$)}}
\rput(2.8,1.7){${\bf r}^{\,\prime}$}

\pscircle[linecolor=black,linewidth=0.8mm](4 , -2){.07}
\rput(4,-1.7){{$-q$}}
\psline[linecolor=black,linewidth=0.2mm]{->}(0,0)(4 , -2)
\rput(5.2,-1.9){{($x',y', -z' $)}}
\rput(2.8,-1.8){${\bf r}^{\,\prime}_i$}

\end{pspicture}
\end{center}
\caption{Point charge $q$ near an infinite conducting plane and its image.}
\label{Atom-Plane}
\end{figure}

Consider a polarizable atom at position $\mathbf{r}_0$ in the presence of an infinite conducting plane located at $z=0$. As outlined
in last section, all we need to obtain the dispersion van der Waals interaction energy for such a system
is to find out the function $G_H({\bf r},{\bf r}^{\,\prime})$ associated to it. FIG. \ref{Atom-Plane} shows the charge $q$ at position
 ${\bf r}^{\,\prime} = (x^{\,\prime},y^{\,\prime},z^{\,\prime})$ and the conducting plane, as well as the image charge $q_i$ which, in this case, is simply given by $q_i = -q$ and is located at position ${\bf r}^{\,\prime}_i = (x^{\,\prime},y^{\,\prime},-z^{\,\prime})$.

The electrostatic potential at $\mathbf{r}$ created by the image charge $-q$ located at position ${\bf r}^{\,\prime}_i$ is
\begin{equation}
\phi_i(\mathbf{r})=\frac{-q}{4\pi\varepsilon_0|\mathbf{r}-\mathbf{r}_i'|} \, ,
\end{equation}
where ${\bf r}^{\,\prime}_i = {\bf r}^{\,\prime} - 2z^{\,\prime}\hat{\bf z}$. Hence, from equation (\ref{imgh}), we readly obtain
\begin{equation}
G_H(\mathbf{r},\mathbf{r}')=\frac{\phi_i(\mathbf{r})}{q\varepsilon_0}=-\frac{1}{4\pi|\mathbf{r}-\mathbf{r}_i'|} \, .
\end{equation}

Now we are ready to use Eberlein and Zietal's method. Substituting the previous expression for $G_H(\mathbf{r},\mathbf{r}')$
into equation (\ref{eberlein}), we have
\begin{equation}\label{UNR}
U\!\!_{ap}({\bf r}_0) =  - \frac{1}{8\pi\varepsilon_0}\sum_{m=1}^3\langle d_m^2\rangle\partial_m\partial_m^{\,\prime}
\left\{\frac{1}{[(x-x')^2+(y-y')^2 + (z+z')^2]^{1/2}}\right\}\Big|_{{\bf r} = {\bf r}^{\,\prime} = {\bf r}_0} \; .
\end{equation}
The derivatives in the previous expression can be easily computed. For instance, for the first coordinate, $m=x$, we get
\begin{equation}
\partial_x\partial_x^{\,\prime}
\Bigl\{\frac{1}{[(x-x')^2+(y-y')^2 + (z+z')^2]^{1/2}}\Bigr\}\Big|_{{\bf r} = {\bf r}^{\,\prime} = {\bf r}_0} =
\; \frac{1}{8 \vert z_0\vert^3}\, .
\end{equation}
An identical result is valid for $m=y$, while for $m=z$, the following result is obtained
\begin{equation}
\partial_z\partial_z^{\,\prime}
\Bigl\{\frac{1}{[(x-x')^2+(y-y')^2 + (z+z')^2]^{1/2}}\Bigr\}\Big|_{{\bf r} = {\bf r}^{\,\prime} = {\bf r}_0} =
\; \frac{1}{4 \vert z_0\vert^3}\, .
\end{equation}
Substituting the previous results into equation (\ref{UNR}), we finally obtain
\begin{equation}\label{atomoplano}
U\!\!_{ap}(z_0) =
-\frac{\langle d_x^2\rangle+\langle d_y^2\rangle+2\langle d_z^2\rangle}{64\pi\varepsilon_0|z_0|^3} \, .
\end{equation}
This is the well-known  interaction between an atom and an infinite conducting plane in
the non-retarded regime firstly obtained in 1932 by Lennard-Jones \cite{Lennard-Jones-1932} (see also \cite{CohenEtAl-1973,Dalibard-2002}).

\subsection{Atom close to a grounded sphere\label{atesf}}

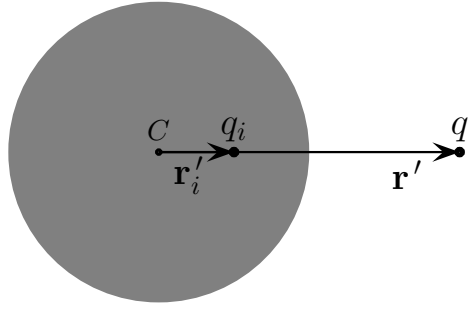
\begin{figure}[!h]
\begin{center}
\newpsobject{showgrid}{psgrid}{subgriddiv=1,griddots=10,gridlabels=6pt}
\begin{pspicture}(-4.5,-1.2)(4.5,2.0)
\psset{arrowsize=0.2 2}
\psset{unit=1}
%
\pscircle[linecolor=gray,linewidth=2cm](0,0){2}

\pscircle[linecolor=black,linewidth=0.4mm](0,0){.05}
\rput(0,0.3){{$C$}}

\pscircle[linecolor=black,linewidth=0.8mm](4,0){.07}
\rput(4,0.3){\large{$q$}}
\rput(3.3,-0.3){\large{${\bf r}^{\,\prime}$}}
\psline[linecolor=black,linewidth=0.3mm]{->}(0,0)(4,0)

\pscircle[linecolor=black,linewidth=0.8mm](1,0){.07}
\rput(1,0.3){\large{$q_i$}}
\rput(0.4,-0.3){\large{${\bf r}^{\,\prime}_i$}}
\psline[linecolor=black,linewidth=0.3mm]{->}(0,0)(1,0)

\end{pspicture}
\end{center}
\caption{Point charge $q$ near a conducting sphere of radius $R$ and its image  $q_i$.}
\label{atesat}
\end{figure}

 Now let us consider an atom in the presence of conducting grounded sphere of radius $R$ and center $C$. The corresponding electrostatic
 problem we need to solve is that of a point charge $q$ at position $\mathbf{r}'=(x',y',z')$ in the presence of the conducting sphere.
The image method for this problem tells us (see, for instance, Griffith's textbook\cite{Griffiths1999}) that we have to put an image
 charge $q_{i} = -\frac{R}{r^{\,\prime}}q$ at position $\mathbf{r}^{\,\prime}_{i} = \frac{R^2}{r'^2}\mathbf{r}'$, where $r^{\,\prime} = \vert{\bf r}^{\,\prime}\vert$, as  sketched in FIG. \ref{atesat}.

\noindent
Hence, the potential of created by the image charge at a generic point $\mathbf{r}$ ($r>R$) is given by
\begin{equation}	
 \phi_i(\mathbf{r}) = \frac{q_i}{4\pi\varepsilon_0|\mathbf{r}-\mathbf{r}^{\,\prime}_{i}|}
 = -\frac{qR}{4\pi\varepsilon_0 r^{\,\prime}|\mathbf{r}-\mathbf{r}^{\,\prime}_{i}|} \, .
\end{equation}
 Substituting this expression into equation (\ref{imgh}) we get the homogeneous solution $G_H(\mathbf{r},\mathbf{r}')$ for this configuration,
\begin{equation}
G_H(\mathbf{r},\mathbf{r}')=\frac{\varepsilon_0\phi_i(\mathbf{r})}{q}
= -\frac{R}{4\pi r^{\,\prime}|\mathbf{r}-\mathbf{r}^{\,\prime}_{i}|} \, .
\end{equation}
Hence, all we have to do is to apply Eberlein-Zietal formula, {\it i.e.}, equation (\ref{eberlein}). For simplicity, we shall consider an isotropic atom, a hypothesis which allows us to write
\begin{equation}
\label{isotr}
\langle d_x^2\rangle = \langle d_y^2\rangle = \langle d_z^2\rangle=\frac{\langle {\bf d}^2\rangle}{3} \, .
\end{equation}
 For convenience, we can orient the axis so that the atom is located at $\mathbf{r}_0=(0,0,z_0)$. Note that with this assumption there is no loss of generalization, since the problem exhibits spherical symmetry. The calculation is lengthier than the previous one for the atom-plane configuration but it still involves only elementary derivatives, so that we shall show only the main steps, leaving for the readers the verification of the intermediate ones.

Let us compute $\partial_x^{\,\prime}\partial_x \Bigl( r^{\,\prime}\vert{\bf r} -
{\bf r}^{\,\prime}_i\vert\Bigr)^{-1}\big|_{{\bf r} = {\bf r}^{\,\prime} = (0,0,z_0)}$\, where
\begin{eqnarray}
 r^{\,\prime} &=& \Bigl[ x^{\,\prime 2} + y^{\,\prime 2} + z^{\,\prime 2}\Bigr]^{1/2}\cr
 \vert{\bf r} - {\bf r}^{\,\prime}_i\vert
  &=&
  \left[\left( x - \frac{R^2}{r^{\,\prime\, 2}}\,x^{\,\prime}\right)^2 +
  \left( y - \frac{R^2}{r^{\,\prime\, 2}}\,y^{\,\prime}\right)^2 +
  \left( z - \frac{R^2}{r^{\,\prime\, 2}}\,z^{\,\prime}\right)^2\right]^{1/2} \; .
\end{eqnarray}
 With this purpose, note initially that
\begin{eqnarray}
\partial_x\left(\frac{1}{r^{\,\prime}\vert{\bf r} - {\bf r}^{\,\prime}_i\vert}\right) =
-\frac{1}{r^{\,\prime}} \frac{\left( x - \frac{R^2}{r^{\,\prime\, 2}}\,x^{\,\prime}\right)}
{\left[\left( x - \frac{R^2}{r^{\,\prime\, 2}}\,x^{\,\prime}\right)^2 +
  \left( y - \frac{R^2}{r^{\,\prime\, 2}}\,y^{\,\prime}\right)^2 +
  \left( z - \frac{R^2}{r^{\,\prime\, 2}}\,z^{\,\prime}\right)^2\right]^{3/2}}
\end{eqnarray}
By computing the $\partial_x^{\,\prime}$ derivative of the previous expression and after that
evaluating the result at ${\bf r}={\bf r}^{\,\prime} = (0,0,z_0)$, it is straightforward to show that
\begin{equation}
\partial_x^{\,\prime}\partial_x\left(\frac{1}{r^{\,\prime}\vert{\bf r} -
 {\bf r}^{\,\prime}_i\vert}\right)\bigg|_{{\bf r} = {\bf r}^{\,\prime} = (0,0,z_0)} =
 \frac{R^2}{z_0^6\left( 1 - R^2/z_0^2\right)^3}
\end{equation}
An identical result is obtained for the coordinate $m=y$. In the case of coordinate $m=z$, it can be shown with an analogous but lengthier calculation that
\begin{equation}
\partial_z^{\,\prime}\partial_z\left(\frac{1}{r^{\,\prime}\vert{\bf r} -
 {\bf r}^{\,\prime}_i\vert}\right)\bigg|_{{\bf r} = {\bf r}^{\,\prime} = (0,0,z_0)} =
 \frac{2R^2}{z_0^6\left( 1 - R^2/z_0^2\right)^3}\; + \;
 \frac{1}{z_0^4\left( 1 - R^2/z_0^2\right)^3}\; .
\end{equation}
Collecting all the previous results and substituting them into (\ref{eberlein}), we finally obtain the dispersion van der Waals interaction energy between an atom and a grounded conducting sphere,
\begin{equation}\label{AtomoEsfera-ZoR}
U\!_{ags}(z_0,R) = -\frac{\langle {\bf d}^2\rangle}{24\pi\varepsilon_0}
 \left\{ \frac{4R^3}{z_0^6}\frac{1}{(1-R^2/z^2_0)^3}+\frac{R}{z_0^4}\frac{1}{(1-R^2/z^2_0)^2} \right\}\, .
\end{equation}
For later convenience, we rewrite the previous result in terms of $R$ and the distance between the atom and the sphere, $z_0 - R$, which we shall denote by $a$. Substituting $a = z_0 -R$ in (\ref{AtomoEsfera-ZoR}) we get, after a few rearrangements,
\begin{equation}\label{AtomoEsfera-aR}
U\!_{ags}(a,R) = -\frac{\langle {\bf d}^2\rangle}{24\pi\varepsilon_0\, a^3}
 \left\{ \frac{4}{(2 + a/R)^3} +  \frac{a/R}{(2 + a/R)^2} \right\}\, .
\end{equation}
For an atom with a dominant transition frequency, we may still cast our result in terms of the static polarizability of the atom. Recalling that the static polarizability for an atom in its ground state is given by \cite{Davydov-Book-1976}
\begin{equation}\label{alpha}
\alpha = \frac{2}{3\hbar} \sum_{n\ne 0} \frac{\vert{\bf d}_{n0}\vert^2}{\omega_{n0}}\, ,
\end{equation}
where $\omega_{n0}$ is the transition frequency between the $n$-th state and the ground state, and ${\bf d}_{n0}$ is the corresponding transition dipole moment, for an atom with a dominat transition, say between the fundamental state and the first excited one, we have
\begin{equation}
\alpha = \frac{2\vert{\bf d}_{10}\vert^2}{3\hbar\omega_{10}}
 \;\;\;\;\Longrightarrow\;\;\;\;
 \vert{\bf d}_{10}\vert^2 = \frac{3\hbar\omega_{10}}{2}\alpha
\end{equation}
Since for an atom with this dominant transition  $\langle {\bf d}^2\rangle = \vert{\bf d}_{10}\vert^2$, equation (\ref{AtomoEsfera-aR}) reduces to
\begin{equation}\label{AtomoEsfera-aR-alpha}
U\!_{ags}(a,R) = -\frac{\hbar\omega_{10}\alpha}{16\pi\varepsilon_0\, a^3}
 \left\{ \frac{4}{(2 + a/R)^3} +  \frac{a/R}{(2 + a/R)^2} \right\}\, .
\end{equation}
Some comments are in order: {\it (i)} the previous result was obtained by the first time by Taddei and collaborators \cite{Taddei-Mendes-Farina-2010}. However, the agreement of our result, given by equation (\ref{AtomoEsfera-aR-alpha}), and the  result  obtained by these authors is up to a numerical factor of $3$. A discrepancy of a numerical factor between both results was expected, since in Ref. \cite{Taddei-Mendes-Farina-2010} the authors employed the (semiclassical) fluctuating-dipoles method, which is not expected to provide the correct numerical factors, though it gives the correct behavior of the interaction;
 {\it (ii)} the atom-sphere system had been discussed before by many authors \cite{Marvin-Toigo-1982,Jhe-Kim-1995A,Jhe-Kim-1995B,BuhmannEtAl-2004} for spheres with different properties and for regimes other than the non-retarded one and recently has been a subject of great interest \cite{Sambale-Buhmann-Scheel-2010,EllingsenEtAl-2011,EllingsenEtAl-2012}. In fact, the result expressed in equation (\ref{AtomoEsfera-aR}) was reobtained by Buhmann as a particular case of a more general discussion \cite{Buhmann-PrivateComunication};
  {\it (iii)} equation (\ref{AtomoEsfera-aR-alpha}) is valid for any values of $R$ and $a$, provided the conditions for the non-retarded regime remain valid. It is easy to show that, in the limit  $R\rightarrow\infty$, with finite  $a$, equation (\ref{AtomoEsfera-aR}) reproduces the non-retarded interaction energy for the atom-plane system, namely,
$U\!_{ags}(z_0) \rightarrow -\langle {\bf d}^2\rangle/[48\pi\varepsilon_0\, a^3]$, in agreement with equation (\ref{atomoplano}) if there we write
$\langle d_x^2\rangle + \langle d_y^2\rangle + 2\langle d_z^2\rangle = (4/3)\langle {\bf d}^2\rangle$.

\subsection{Atom close to an isolated conducting sphere}

Let us consider now with a neutral perfectly conducting isolated sphere. This case differs from that of an atom close to a grounded sphere because now $G_H({\bf r},{\bf r}^{\,\prime})$ does not satisfy the boundary condition written in (\ref{ccgat}). Since the sphere is not grounded anymore, the presence of a dipole changes its potential. Hence, the first thing we have to do is to find out the electrostatic potential on the surface or the sphere. With the aid of the image method, it is possible to show that the potential induced on an isolated sphere by a dipole at position $\mathbf{r}_0$ is given by \cite{Santos-Tort-2004}
\begin{equation}
V_{sph} = \frac{\mathbf{d}\cdot\mathbf{r}_0}{4\pi\varepsilon_0|\mathbf{r}_0|^3} \, .
\end{equation}
Therefore, the BC to be satisfied by the electrostatic potential in this problem is
\begin{equation}
\Phi(\mathbf{r})\big|_{|\mathbf{r}|=R}=\frac{\mathbf{d}\cdot\mathbf{r}_0}{4\pi\varepsilon_0|\mathbf{r}_0|^3} \,. \label{ccphi}
\end{equation}
Substituting the above result  into equation (\ref{phig}), we obtain
\begin{equation}
\int\!\! G(R \,\hat{\mathbf{r}},\mathbf{r}')\rho(\mathbf{r}')\, d^3\mathbf{r}'
= \frac{\mathbf{d}\cdot\mathbf{r}_0}{4\pi|\mathbf{r}_0|^3} \, .
\end{equation}
Using the same charge density as before, namely, that given by (\ref{denscargaeb}), and following the same procedure employed to go from equation (\ref{taylor1}) to equation (\ref{taylor2}), we get
\begin{equation}\label{ccesfate}
	\mathbf{d}\cdot\nabla^{\,\prime} G(R\, \hat{\mathbf{r}},\mathbf{r}')\big|_{\mathbf{r}'=\mathbf{r}_0}=\frac{\mathbf{d}\cdot\mathbf{r}_0}{4\pi|\mathbf{r}_0|^3} \, .
\end{equation}
Hence, in order to the boundary condition (\ref{ccphi}) be fulfilled, we may impose the following BC to the Green function,
\begin{equation}\label{ccghis}
	\nabla'G(\mathbf{r},\mathbf{r}')\bigg|_{{\bf r} \, = \,R\,\hat{\mathbf{r}}
 \atop{{\bf r}^{\,\prime} = \,\mathbf{r}_0}} = \frac{\mathbf{r}_0}{4\pi|\mathbf{r}_0|^3} \, .
\end{equation}
Apart from this condition, the  Green function must also obey Poisson equation (\ref{green}).
As it will become evident in a moment, a convenient  way to obtain the desired Green function is write it as the sum of the Green function of the previous case (with the grounded sphere), with an extra term  $G^{(1)}({\bf r},{\bf r}^{\,\prime})$ to be determined,
\begin{equation}\label{ghcasois}
G(\mathbf{r},\mathbf{r}') = \frac{1}{4\pi|\mathbf{r}-\mathbf{r}'|} - \frac{R}{4\pi|\mathbf{r}'||\mathbf{r}-{\bf r}^{\,\prime}_{i}|} \;+\; G^{(1)}(\mathbf{r},\mathbf{r}') \, .
 \end{equation}
The first two terms on the right hand side of the previous equation, together, satisfy
the Poisson equation in the presence of a grounded sphere, so that these two terms (when added), vanish on the surface of the conductor.
Therefore, in order to solve the problem with an isolated sphere, we just impose that $G^{(1)}({\bf r},{\bf r}^{\,\prime})$ must satisfy the Laplace equation as well as the boundary condition (\ref{ccghis}), to wit,
 \begin{eqnarray}
	\nabla^2 G^{(1)}(\mathbf{r},\mathbf{r}')
&=&
0 \label{lapgis} \\
 \nabla^{\,\prime} G^{(1)}(\mathbf{r},\mathbf{r}')\bigg|_{{\bf r} \, = \,R\,\hat{\mathbf{r}}
 \atop{{\bf r}^{\,\prime} = \,\mathbf{r}_0}}
 &=&
 \frac{\mathbf{r}_0}{4\pi|\mathbf{r}_0|^3} \, . \label{ccgis}
\end{eqnarray}
Except for a constant factor, the right hand side of (\ref{ccgis}) is identified with the electric field created by a point charge
at the origin. Since the electromagnetic is minus the gradient of the electrostatic potential,
$G^{(1)}(R\mathbf{\hat{r}},\mathbf{r}')$  is naturally identified with the electrostatic potential created at a point $\mathbf{r}'$
by a point charge at the origin. Note that this is compatible with equation (\ref{lapgis}), since this
equation must be satisfied only in the physical region of the problem at hand, namely, the region outside
the sphere. Hence, we may write
\begin{equation}
	G^{(1)}(\mathbf{r},\mathbf{r}')=\frac{f(\mathbf{r})}{4\pi|\mathbf{r}'|} \, ,
\end{equation}
where $f(\mathbf{r})$  must assume the unit value
for points belonging to the surface of the sphere, in order to satisfy equation (\ref{ccgis}).
Using the fact that the  Green function must be
symmetric \cite{Arfken1985} by the exchange $\mathbf{r}\longleftrightarrow\mathbf{r}'$, and since $f(\mathbf{r})=1$ for
$|\mathbf{r}| = R$, we are led to the result
\begin{equation}\label{G1}
	G^{(1)}(\mathbf{r},\mathbf{r}')=\frac{R}{4\pi|\mathbf{r}||\mathbf{r}'|} \, .
\end{equation}
Inserting (\ref{G1}) into (\ref{ghcasois}) we see that the $G_H(\mathbf{r},\mathbf{r}')$ function for the present
case is given by
\begin{eqnarray}
	G_H(\mathbf{r},\mathbf{r}') &=& G({\bf r},{\bf r}^{\,\prime}) - \frac{1}{4\pi|\mathbf{r}-\mathbf{r}'|}\cr
&=&
-\frac{R}{4\pi|\mathbf{r}'||\mathbf{r}-\mathbf{r}^{\,\prime}_{i}|}+\frac{R}{4\pi|\mathbf{r}||\mathbf{r}'|} \, ,
\end{eqnarray}
where $\mathbf{r}^{\,\prime}_i=\frac{R^2}{r'^2}\mathbf{r}'$, as in the grounded case.

As we have seen, some care must be taken with problems involving isolated conductors, because
the boundary condition will depend on the charge distribution of the system.
%
%
We should also emphasize the convenience of having used the solution of the problem with a grounded sphere as a step in the
search of a solution to the problem with an isolated sphere.

 As one may readily verify, equations (\ref{energyg})-(\ref{eberlein}) do not depend on the boundary conditions satisfied by $G_H({\bf r},{\bf r}^{\,\prime})$ and, as a consequence, Eberlein-Zietal expression (\ref{eberlein}) is still valid in this case. Using the previous results for the grounded sphere, as well as  the following relations
involving $G^{(1)}({\bf r},{\bf r}^{\,\prime})$,
\begin{eqnarray}
\partial_x\partial_x^{\,\prime}
 \left(\frac{R}{4\pi\vert{\bf r}\vert\vert{\bf r}^{\,\prime}\vert}\right)\bigg|_{{\bf r} = {\bf r}^{\,\prime} = (0,0,z_0)} \;& =&\;
 \partial_y\partial_y^{\,\prime}
 \left(\frac{R}{4\pi\vert{\bf r}\vert\vert{\bf r}^{\,\prime}\vert}\right)\bigg|_{{\bf r} = {\bf r}^{\,\prime} = (0,0,z_0)} \; =0\cr\cr
 \partial_z\partial_z^{\,\prime}
 \left(\frac{R}{4\pi\vert{\bf r}\vert\vert{\bf r}^{\,\prime}\vert}\right)\bigg|_{{\bf r} = {\bf r}^{\,\prime} = (0,0,z_0)}\;
 &=&
 \frac{R}{4\pi z_0^4}
\end{eqnarray}
we finally obtain the dispersion van der Waals interaction energy between an atom and an isolated conducting sphere,
\begin{equation}\label{atomesfis}
	U\!_{ais}(z_0,R) = -\frac{\langle d^2\rangle}{24\pi\varepsilon_0} \left\{ \frac{4R^3}{z_0^6}\frac{1}{(1-R^2/z^2_0)^3}+\frac{R}{z_0^4}\frac{1}{(1-R^2/z^2_0)^2}-\frac{R}{z_0^4} \right\}\, .
\end{equation}
As in the previous section, we can also write last expression in terms of $R$ and the distance from the atom to the surface of the sphere, $z_0 -R$, denoted by $a$. Doing this, we have
\begin{equation}\label{AtomoEsferaIsolada-aR}
U\!_{ais}(a,R) = -\frac{\langle {\bf d}^2\rangle}{24\pi\varepsilon_0\, a^3}
 \left\{ \frac{4}{(2 + a/R)^3} +  \frac{a/R}{(2 + a/R)^2} - \frac{a^3/R^3}{(1 + a/R)^4}\right\}\, ,
\end{equation}
in agreement, up to a numerical factor of $3$, with the result obtained by Taddei and collaborators \cite{Taddei-Mendes-Farina-2010}.
The second term on the right hand side of last equation is, in absolute value, greater than the third. Therefore,
the interaction of an atom with an isolated conducting sphere is always attractive. Since the only difference between the grounded and isolated
cases is the last term present in equation (\ref{atomesfis}), we conclude that the attraction is stronger in the
 case of a grounded sphere. This can be physically understood as a consequence of the charge acquired by the grounded sphere.

We finish this section by taken the interesting limit $R\rightarrow 0$, but with $4\pi\varepsilon_0R^3 \rightarrow \alpha_s$, where $\alpha_s$ is the (finite) polarizability of a very small conducting sphere. In this limit, the previous equation reduces to
\begin{equation}\label{PontoCondutor}
\lim_{R\rightarrow 0\atop{\alpha_s\, =\, 4\pi\varepsilon_0 R^3 }}
U\!_{ais}(a,R) = -\frac{\langle {\bf d}^2\rangle}{24\pi\varepsilon_0\, a^3}
 \frac{4R^3}{a^3} \; =\;
 -\, \frac{\hbar\omega_{10}\alpha\alpha_s}{(4\pi\varepsilon_0)^2\, a^6}\, ,
\end{equation}
where in the last step we assumed that the transition from the fundamental state to the first excited one is dominant. Note that the result is a London like dipole-dipole interaction, as expected.

\section{Atom close to a conducting boss hat surface}

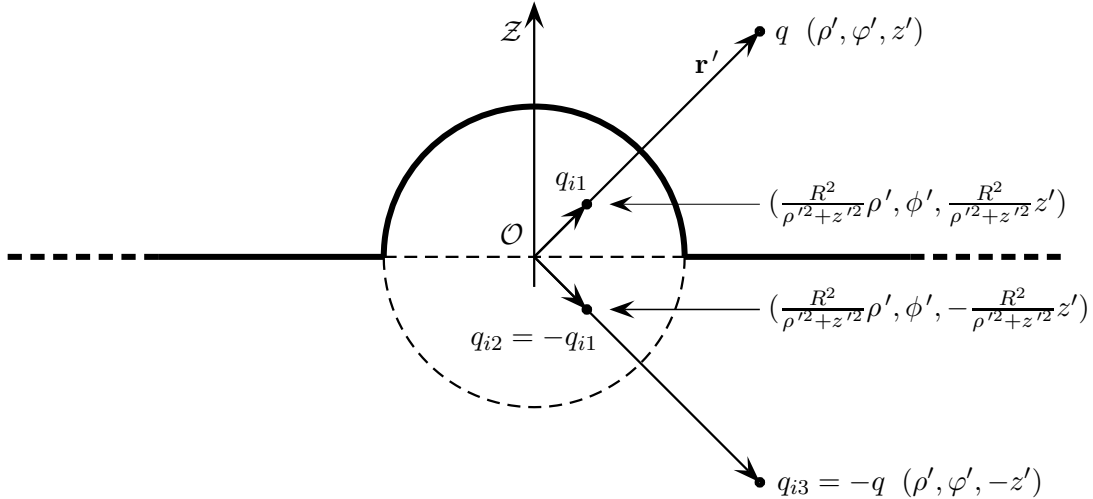
\begin{figure}[!h]
\begin{center}
\newpsobject{showgrid}{psgrid}{subgriddiv=1,griddots=10,gridlabels=6pt}
\begin{pspicture}(-4.5,-2.2)(5.5,3.0)
\psset{arrowsize=0.2 2}
\psset{unit=1}
%
\psline[linecolor=black,linewidth=0.3mm]{->}(0,-0.4)(0,3.4)
\rput(-0.3,0.3){${\cal O}$}
\rput(-0.3,3.0){${\cal Z}$}

\psline[linecolor=black,linewidth=0.8mm,linestyle=dashed]{-}(-7,0)(-5 ,0)
\psline[linecolor=black,linewidth=0.8mm]{-}(-5,0)(-2 ,0)
\psarc[linecolor=black,linewidth=0.8mm]{-}(0,0){2}{0}{180}
\psarc[linecolor=black,linewidth=0.3mm,linestyle=dashed]{-}(0,0){2}{180}{360}
\psline[linecolor=black,linewidth=0.8mm]{-}(2,0)(5,0)
\psline[linecolor=black,linewidth=0.8mm,linestyle=dashed]{-}(5,0)(7 ,0)
\psline[linecolor=black,linewidth=0.3mm,linestyle=dashed]{-}(-2,0)(2 ,0)
%
\pscircle[linecolor=black,linewidth=0.8mm](3,3){.07}
\rput(4.2,3.0){$q\;\;(\rho ',\varphi ', z')$}
\psline[linecolor=black,linewidth=0.3mm]{->}(0,0)(3,3)
\rput(2.3,2.6){${\bf r}^{\,\prime}$}

\pscircle[linecolor=black,linewidth=0.8mm](0.7 , 0.7){.07}
\rput(0.5,1.0){{$q_{i1}$}}
\psline[linecolor=black,linewidth=0.3mm]{->}(0,0)(0.7 , 0.7)
\psline[linecolor=black,linewidth=0.1mm]{<-}(1.0,0.7)(3.0,0.7)
\rput(5.1,0.7){$(\frac{R^2}{\rho^{\,\prime 2} + z^{\,\prime 2}}\rho^{\,\prime},\phi^{\,\prime},
 \frac{R^2}{\rho^{\,\prime 2} + z^{\,\prime 2}}z')$}
\pscircle[linecolor=black,linewidth=0.8mm](3,-3){.07}
\psline[linecolor=black,linewidth=0.3mm]{->}(0,0)(3,-3)
\rput(5,-3){$q_{i3} = -q \;\; (\rho ',\varphi ', -z')$}
\pscircle[linecolor=black,linewidth=0.8mm](0.7 , -0.7){.07}
\rput(0,-1.1){{$q_{i2} = -q_{i1}$}}
\psline[linecolor=black,linewidth=0.3mm]{->}(0,0)(0.7 , -0.7)
\psline[linecolor=black,linewidth=0.2mm]{<-}(1.0,-0.7)(3.0,-0.7)
\rput(5.25,-0.7){$(\frac{R^2}{\rho^{\,\prime 2} + z^{\,\prime 2}}\rho^{\,\prime},\phi^{\,\prime},
 -\frac{R^2}{\rho^{\,\prime 2} + z^{\,\prime 2}}z')$}

\end{pspicture}
\end{center}
\caption{A physical point charge $q$ close to the conducting boss hat surface and the three necessary image charges located in the non-physical region.}
\label{atombosshat}
\end{figure}

Having solved the simple cases of the last section, we are now in position to solve a more
interesting case, namely, an atom near a the conducting surface with the shape of a \lq\lq boss hat{\rq\rq}. This surface consists of a conducting spherical hemisphere with radius $R$ together with an infinite conducting plane. The geometry in question and the necessary image charges to the problem are sketched in FIG. \ref{atombosshat}.

To begin with, consider a charge $q$ at position ${\bf r}^{\,\prime}$ in the
presence of the conducting boss hat. Now put a charge $q_{i1} = -\frac{R}{r^{\,\prime}}q$ at position
${\bf r}^{\,\prime}_{i1} = \frac{R^2}{r^{\,\prime 2}}\mathbf{r}^{\,\prime}$. In cylindrical coordinates these equations take the form
$ q_{i1} = \frac{R}{\sqrt{\rho^{\,\prime 2} + z^{\,\prime 2}}}q$ and
 ${\bf r}^{\,\prime}_{i1} = (\frac{R^2}{\rho^{\,\prime 2} + z^{\,\prime 2}}\rho^{\,\prime},\phi^{\,\prime},
 \frac{R^2}{\rho^{\,\prime 2} + z^{\,\prime 2}}z')$,
 since $r^{\,\prime} = \vert{\bf r}^{\,\prime}\vert = \sqrt{\rho^{\,\prime 2} + z^{\,\prime 2}}$.
 As used in the atom-sphere case, this pair of charges, $q$ and $q_{i1}$, furnishes a null potential at the spherical part of the
 conducting surface. But the potential generated by these two charges  does not yet
 satisfy the desired BC at the plane part of the conductor. Therefore, we must
introduce two more image charges, one, with charge $q_{i2} = - q_{i1}$, being the mirror image of $q_{i1}$ and the other
with charge $q_{i3} = - q$, being the mirror image of the physical charge $q$. The addition of these two charges, $q_{i2}$ and $q_{i3}$, leads
to a null potential at the plane $z=0$ but with the advantage of not disturbing the null potential at the hemisphere, since by symmetry $q_{i2}$ is precisely the image charge  of $q_{i3}$ with respect to the sphere. Consequently, the four charges, the real charge $q$ plus the three image charges
$q_{i1}$, $q_{i2}$ and $q_{i3}$ lead to an electrostatic potential which is zero on the  boss hat conducting surface. The positions of the four charges, namely, ${\bf r}^{\,\prime} =  (\rho ',\varphi ', z')$,
 ${\bf r}^{\,\prime}_{i1} = (\frac{R^2}{\rho^{\,\prime 2} + z^{\,\prime 2}}\rho^{\,\prime},\phi^{\,\prime},
 \frac{R^2}{\rho^{\,\prime 2} + z^{\,\prime 2}}z')$, ${\bf r}^{\,\prime}_{i2} = (\frac{R^2}{\rho^{\,\prime 2} + z^{\,\prime 2}}\rho^{\,\prime},\phi^{\,\prime},
 -\frac{R^2}{\rho^{\,\prime 2} + z^{\,\prime 2}}z')$ and ${\bf r}^{\,\prime}_{i3} =  (\rho ',\varphi ', -z')$
 are indicated in FIG. \ref{atombosshat}.

With this image configuration, the potential generated by the image charges is just a superposition
of the potentials created by the charges $q_{i1}$, $q_{i2}$ and $q_{i3}$, namely,
\begin{eqnarray} \phi_i(\mathbf{r}) &=&
\frac{1}{4\pi\varepsilon_0}\left\{\frac{q_{i1}}{|{\bf r} - {\bf r}^{\,\prime}_{i1}|} +
\frac{q_{i2}}{|{\bf r} - {\bf r}^{\,\prime}_{i2}|} + \frac{q_{i3}}{|{\bf r} - {\bf r}^{\,\prime}_{i3}|}\right\}\, .
\end{eqnarray}
Using the previous expressions for $q_{i1}$, $q_{i2}$ and $q_{i3}$, as well as for ${\bf r}^{\,\prime}_{i1}$, ${\bf r}^{\,\prime}_{i2}$ and ${\bf r}^{\,\prime}_{i3}$, and defining the quantities
$\xi({\bf r},{\bf r}^{\,\prime})$, $\xi_-({\bf r},{\bf r}^{\,\prime})$ and  $\xi_+({\bf r},{\bf r}^{\,\prime})$ as
\begin{equation}
\xi_-({\bf r},{\bf r}^{\,\prime}) = \vert{\bf r} - {\bf r}^{\,\prime}_{i1}\vert\, ;\;\;\;
\xi_+({\bf r},{\bf r}^{\,\prime}) = \vert{\bf r} - {\bf r}^{\,\prime}_{i2}\vert\, ;\;\;\;
\xi({\bf r},{\bf r}^{\,\prime}) = \vert{\bf r} - {\bf r}^{\,\prime}_{i3}\vert\, ,
\end{equation}
the function $G_H({\bf r},{\bf r}^{\,\prime})$ for the boss hat case can be written in the form
\begin{eqnarray}
G_H({\bf r}, {\bf r}') = \dfrac{1}{4\pi}\left\{-\dfrac{1}{\xi({\bf r},{\bf r}')}
- \dfrac{R\sqrt{\rho'^2+z'^2}}{\xi_-({\bf r},{\bf r}')} + \dfrac{R\sqrt{\rho'^2+z'^2}}{\xi_+({\bf r},{\bf r}')} \right\}\, ,
\end{eqnarray}
where
\begin{eqnarray}
\xi({\bf r},{\bf r}') &=& \sqrt{\rho'^2+\rho^2+(z'+z)^2-2\rho'\rho\cos(\phi'-\phi)}\, , \\
\xi_{\pm}({\bf r},{\bf r}') &=& \sqrt{R^4\rho'^2+(\rho'^2+z'^2)^2\rho^2+[(\rho'^2+z'^2)z\pm
R^2z']^2-2R^2(\rho'^2+z'^2)\rho'\rho\cos(\phi'-\phi)}\ \ \ \ \ \ \
\end{eqnarray}
and we also used equation (\ref{imgh}).

\begin{figure}[!h]
\centering
\includegraphics[scale=0.57]{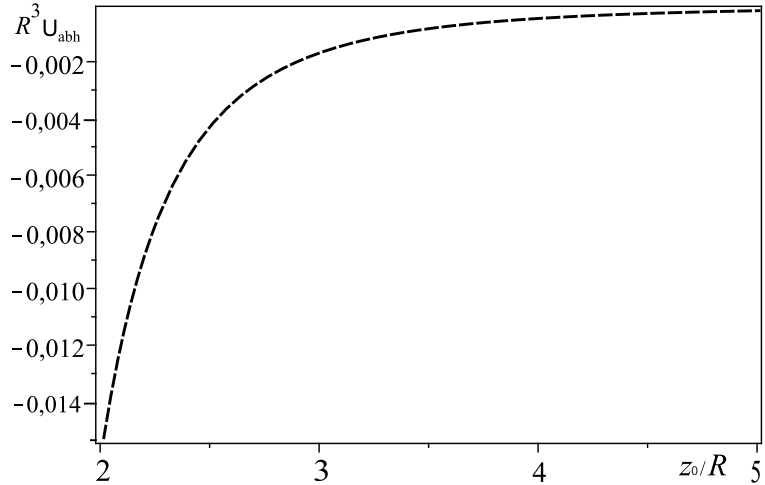}
\caption{Dispersion interaction energy between an atom and a conducting boss hat surface as a function of $z_0/R$ with the atom on the ${\cal OZ}$ axis and assuming that $\langle d_z^2\rangle\gg \langle d_\rho^2\rangle,\, \langle d_\phi^2\rangle$. The graph is plotted in arbitrary units.}
\label{BossHat1}
\end{figure}

The dispersion interaction energy of an atom at a generic position  $(\rho_0,\phi_0,z_0)$ and a boss hat conducting surface, $U\!_{abh}(\rho_0,\phi_0,z_0)$, can then be obtained from equation (\ref{eberlein}). After a lengthy but straightforward calculation, it can be shown that
\begin{equation}\label{Uabh}
U\!_{abh}(\rho_0,z_0)  = -\dfrac{1}{64\pi\varepsilon_0z_0^3}\Bigl\{\langle d^2_{\rho}\rangle \Xi_{\rho}(\rho_0,z_0) +
\langle d^2_{\varphi}\rangle \Xi_{\varphi}(\rho_0,z_0) + \langle d^2_{z}\rangle\Xi_{z}(\rho_0,z_0)\Bigr\}\, ,
\end{equation}
where
\begin{eqnarray}
\Xi_{\rho}(\rho_0,z_0) &=& 1
- 8Rz_0^3\left\{ \dfrac{[(R^2+z_0^2)^2+(R^2-\rho_0^2-8z_0^2)\rho_0^2]R^2 +
(z_0^2 + \rho_0^2)^2\rho_0^2}{[(\rho_0^2 + z_0^2+R^2)^2-4R^2\rho_0^2]^{5/2}}
- \dfrac{\rho_0^2 + R^2}{(\rho_0^2 + z_0^2 - R^2)^3} \right\} \ \ \ \ \ \ \ \ \\\cr
\Xi_{\varphi}(\rho_0,z_0) &=&
1 + 8R^3z_0^3\left\{\dfrac{1}{(\rho_0^2 + z_0^2 - R^2)^3} -
\dfrac{1}{[(\rho_0^2 + z_0^2 + R^2)^2 - 4R^2\rho_0^2]^{3/2}} \right\}  \\\cr
\Xi_{z}(\rho_0,z_0) &=&
2 + \dfrac{8Rz_0^3}{(\rho_0^2 + z_0^2 - R^2)^3}
\left\{R^2 + z_0^2+ \dfrac{\zeta(R,\rho_0,z_0)}{[(\rho_0^2 + z_0^2 + R^2)^2-4R^2\rho_0^2]^{5/2}} \right\}
\end{eqnarray}
with
\begin{eqnarray}
\zeta(R,\rho_0,z_0) = &-&
 R^2\rho_0^2\Bigl[-10\rho_0^4z_0^4 - 10\rho_0^4R^2z_0^2 - 10R^4\rho_0^4 + 8\rho_0^2R^4z_0^2 - z_0^8 + 2\rho_0^6z_0^2+ \cr
                   &+&
                   8\rho_0^2z_0^6 - 36\rho_0^2R^2z_0^4 + 10\rho_0^2R^6\Bigr] - (R^4 - z_0^4)^2(R^2-z_0^2)^2 \cr
                   &-&
                   5\rho_0^2z_0^4(z_0^2 + \rho_0^2)\Bigl[(z_0^2 + \rho_0^2)^2-\rho_0^2z_0^2\Bigr]\, .
\end{eqnarray}
As expected,  the interaction energy does not depend on $\phi$, due to the axial symmetry of the system. Further, one may immediately recover the atom-plane result, given by (\ref{atomoplano}), just taking $R=0$ in the previous equations. In FIG. \ref{BossHat1} we plot the interaction energy given by (\ref{Uabh}) multiplied by $R^3$ (apart from a constant factor) in terms of $z_0/R$ for the particular case where the atom is on the ${\cal OZ}$ axis ($\rho_0 = 0$) and its atomic polarizability in this direction is dominant, {\it i.e.}, $\langle d_z^2\rangle\gg \langle d_\rho^2\rangle ,\, \langle d_\phi^2\rangle$. Note that only the function $\Xi_{z}$ is necessary.


\begin{figure}[!h]
\centering
\includegraphics[scale=0.48]{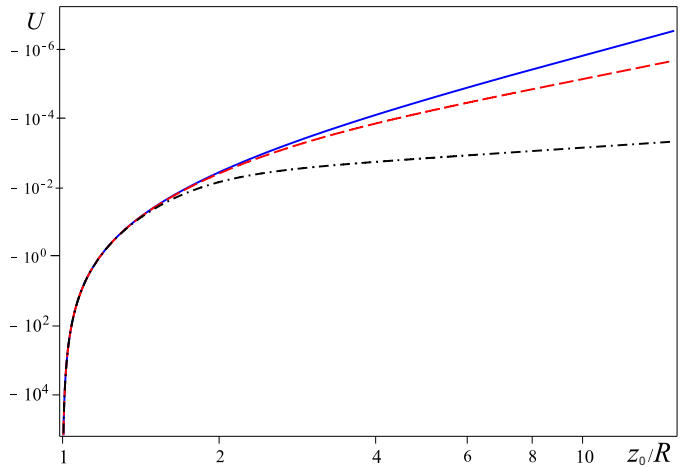}
\caption{(Color online) Exact expression for $U\!_{ags}$ (blue solid line), given by (\ref{AtomoEsfera-ZoR}) and  expansions up to third order for  $U\!_{abh}$ (dashed red line) and $U\!_{ags}$ (dotted-dashed black line), given respectively by (\ref{U3ags}) and (\ref{U3abh}), as functions of $z_0/R$. Both axis are in log scale and we are using arbitrary units.}
\label{Comparacao3aOrdem}
\end{figure}

It is interesting to analyze the curvature effects on the interaction between the atom and the boss hat surface by comparing the interaction for this case, given by   (\ref{Uabh}) with that for the atom-grounded sphere case, given by (\ref{AtomoEsfera-ZoR}). To be consistent, we shall now consider an isotropic atom in equation (\ref{Uabh}), since we made this assumption in obtaining (\ref{AtomoEsfera-ZoR}). However, we shall compare these two expressions only up to third order in $(z_0 - R)/R$. Making, then, a Taylor expansion of equations (\ref{AtomoEsfera-ZoR}) and (\ref{Uabh}) and maintaining only terms up to third order, we obtain for the respective expressions of $U\!_{ags}(z_0,R)$ and $U\!_{abh}(z_0,R)$:
\begin{equation}\label{U3ags}
U\!_{ags}(z_0,R) =  - \frac{\langle{\bf d^2}\rangle}{48 \pi \varepsilon_0(z_0 - R)^3}
\left\{ 1 - \frac{z_0 - R}{R} + \frac{(z_0 - R)^2}{R^2} - \frac{7(z_0 - R)^3}{8R^3}\; +\; ...\right\}
%
\end{equation}
and
\begin{equation}\label{U3abh}
U\!_{abh}(z_0,R) = - \frac{\langle{\bf d^2}\rangle}{48 \pi \varepsilon_0(z_0 - R)^3}
\left\{ 1 - \frac{z_0 - R}{R} + \frac{(z_0 - R)^2}{R^2} - \frac{7(z_0 - R)^3}{8R^3}\; +\; ...\right\}
%
\end{equation}
Comparing last equations we see that they coincide up to second order in $(z_0 - R)/R$ (up to
 order  $(z_0 - R)^2/R^2$, the interaction of an atom with a boss hat surface is the same as that of an atom with a sphere). This is reasonable, since an atom very close to a boss hat surface can not distinguish it from a sphere. As the distance between the atom and the boss hat increases, the differences between the two surfaces become apparent. The comparison between  $U\!_{ags}(z_0,R)$ and $U\!_{abh}(z_0,R)$ expanded up to third order in $(z_0 - R)/R$ and the exact result for the interaction in the atom-grounded sphere case is illustrated in FIG. \ref{Comparacao3aOrdem}.


\section{Conclusions and final remarks}

We discussed here a  method for computing the van der Waals interaction between an atom and a  conducting surface of arbitrary shape introduced recently by Eberlein and Zietal \cite{Eberlein-Zietal-2007} and applied it in a variety of situations, from introductory examples, as the cases of an atom interacting with an infinitely grounded conducting plane, or with a conducting sphere (grounded or isolated), to the more complex case of an atom interacting with a conducting boss hat surface. This method has the advantage of
quickly leading us to a corresponding classical problem in electrostatics, which allows us to use well establihed methods for solving Laplace or Poisson equations as, for instance,  those based on Green functions or on the image method. Particularly, we have shown  explicitly how the image method can be appropriately employed jointly with Eberlein and Zietal's method (when the problem admits an image solution).
We should emphasize, however, that the use of the image method is not mandatory. In fact, in the pioneering work on this method  \cite{Eberlein-Zietal-2007}, the authors discussed the problems of an atom interacting with an infinite conducting semi-plane, and an atom interacting with a conducting cylinder using the Green function method. Also, in further papers, this method was employed without the use of the image method, as in the calculation of the non-retarded interaction between an atom and a dielectric slab \cite{Contreras-Eberlein-2009} and in the computation of the non-retarded interaction of an atom and an infinitely conducting plane with a circular hole \cite{Eberlein-Zietal-2011} (this same system as well as the atom-disk system were discussed with the aid of the more involved Sommerfeld's image method \cite{Sommerfeld-1896} in a recent article \cite{ReinaldoEtAl-2011}). Hence, it is not an abuse to say that Eberlein and Zietal's method  is  due to its simplicity and at the same time its power, with a very favorable cost-benefit analysis for undergraduate and also graduate students that are beginning to study non-retarded dispersion forces between atoms and conducting surfaces. A lot of other systems can be handled by this method. We challenge the interested reader, for instance, to reobtain the non-retarded force between an atom and a conducting wedge with aperture angle equal to $\pi/n$, with $n$ a positive integer, first obtained by Mendes and collaborators \cite{MendesEtAl-2008} (the retarded Casimir-Polder interaction between an atom and a conducting wedge had already been calculated by Brevik and collaborators \cite{BrevikEtAl-1998}).

There are many other methods of computing dispersion forces between atoms and macroscopic bodies which are much more general than the one discussed here, in the sense that they consider all the distance regimes, thermal effects as well as all kinds of materials, not only perfectly conducting ones. It is worth mentioning, for instance, the famous results obtained by Lifshitz in 1956 \cite{Lifshitz-1956}, and generalized by Dzyaloshinskii, Lifshitz and Pitaevskii \cite{DLP-1961} a couple of years later, that since then have been applied with an enormous success. Their results
predicted, for instance, the variation of the thickness of thin superfluid helium films in a very good
agreement with the experiments \cite{Sabisky-Anderson-1973}. However, the discussion of more general methods of computing dispersion forces would lead us very far from the main purposes of this pedagogical article.

We finish this article by mentioning that dispersive forces are still the subject of an intense research, mainly in connection with repulsive dispersive forces, which would be very important in a variety of situations \cite{LevinEtAl-2010,Eberlein-Zietal-2011,ReinaldoEtAl-2011,McCauleyETAl-2011,Rodriguez-Lopez-2011,BostromEtAl-2012,Bostrom-Sernelius-2012,Shajesh-Shaden-2012}, to mention just a few recent results on repulsive forces. We should mention that Feinberg and Sucher \cite{Feinberg-Sucher-1968,Feinberg-Sucher-1970} had already shown that an electrically polarizable atom and a magnetically polarizable one repelled each other. This result has been reotained in a simple way in Ref(s) \cite{Farina-Santos-Tort-2002AJP,Farina-Santos-Tort-2002} (in connection with these results, see also the papers by Boyer \cite{Boyer-1969,Boyer-1974,Dorota-1993}). It is worth mentioning that the interaction between an electrically polarizable atom and a magnetically polarizable one has been discussed in a more general context, namely, with the atoms embedded in a magneto-dielectric medium by the authors \cite{SpagnoloEtAl-2007,Scheel-Buhmann-2009}. The dispersion interaction between a ground state atom and a corrugated surface was firstly discussed by Messina {\it et al$\,$} \cite{MessinaEtAl-LateralCP-2009}, where the so called scattering approach was employed. The existence of a lateral Casimir and Polder force gave rise to several interesting proposals of experiments on dispersive forces, some of them including Bose-Einstein condensates near periodic gratings. We hope this paper can motivate the readers to turn their attention to such an interesting and interdisciplinary subject as the dispersion forces are.

\begin{acknowledgments}
The authors are indebted to F.S.S. Rosa and P.A. Maia Neto for helpful discussions. C.F. thanks the hospitality of people of Trindade where part of this work was done. The authors would like to thank CNPq for partial financial support.
\end{acknowledgments}


\begin{thebibliography}{99}


\bibitem{Israelachvili-2011} Jacob Israelashivili,  {\it Intermolecular $\&$ Surface Forces}, 3rd ed. (Academic Press, New York, 2011).

\bibitem{Margenau-Kestner-1969} H. Margenau and N.R. Kestner, {\it Theory of Intermolecular Forces} (Pergamon, New York, 1969).

\bibitem{VanDerWaals-1873} J.D. van der Waals, {\it Over de continuiteit van den gas-en vloeistoftostand}, (Dissertation, Leiden, 1873).

\bibitem{Reinganum-1903} Max Reinganum, \lq\lq Über Molekularkräfte und elektrische Ladungen der Molekule{\rq\rq}, Ann. Physik {\bf 10}, 334-353 (1903).

\bibitem{Reinganum-1912} Max Reinganum, \lq\lq Kräfte elektrischer Doppelpunkte  nach der statistischen Mechanik und Anwendung auf molekulare und ionenwirkungen{\rq\rq}, Ann. Physik {\bf 343}, 649-668 (1912).

\bibitem{Keesom-1915} W.H. Keesom, \lq\lq The second virial coefficient for rigid spherical molecules, whose mutual attraction is equivalent to that of a quadruplet placed at its centre {\rq\rq}, Proc. Acad. Sci. Amsterdam, {\bf 18}, 636-646	(1915).

\bibitem{Keesom-1920} W.H. Keesom, \lq\lq Quadrupole moments of the oxygen and nitrogen molecules {\rq\rq}, Proc. Acad. Sci. Amsterdam, {\bf 23}, 939-942 (1920).

\bibitem{Debye-1920} P. Debye, \lq\lq van der Waals' cohesion forces{\rq\rq}, Physik Z. {\bf 21} 178-187 (1920).

\bibitem{Debye-1921} P. Debye, \lq\lq Molecular forces and their electrical interpretation {\rq\rq}, Physik Z. {\bf 22} 302-308 (1921).

\bibitem{Wang-1927} S.C. Wang, \lq\lq {\rq\rq}, Phys. Zeit. {\bf 28}, 663 (1927).

\bibitem{Epstein-1926} P.S. Epstein, \lq\lq The new quantum theory and the Zeeman effect{\rq\rq}, Proc. Natl. Acad. Sci. {\bf 12}, 634-638 (1926).

\bibitem{Epstein-1927} P.S. Epstein, \lq\lq The dielectric constant of atomic hydrogen in  undulatory mechanics.{\rq\rq}, Proc. Natl. Acad. Sci. {\bf 13}, 432-438 (1927).

\bibitem{Eisenschitz-London-1930} R. Eisenschitz and F. London \lq\lq Über das Verhältnis der van der Waalsschen Kräfte
zu den homöopolaren Bindungskräften'', Z.Phys. {\bf 60}, 491-527 (1930).

\bibitem{London-1930} F. London, \lq\lq Zur Theorie und Systematik der Molekularkräfte'', Z.Phys. {\bf 63}, 245-279 (1930).

\bibitem{Bransden-Joachain-2000} B.H. Bransden, C.J. Joachain, {\it Quantum Mechanics} (Benjamin Cummings, Harlow, U.K.,2000), p.719.

\bibitem{CohenEtAl-1973} C.Cohen-Tannoudji, B.Diu and F.Laloë, {\it Mécanique Quantique} (Hermann, Paris, 1973),
 Compl. $C_{XI}$ Tome 2.

\bibitem{Berg-Book-2010} John C. Berg, {\it An Introduction to INTERFACES and COLLOIDS: The Bridge to Nanoscience}
(World Scientific, New Jersey, 2010).

\bibitem{Verwey-1947}  E.J.W. Verwey, \lq\lq Theory of the stability of lyophobic collids. {\rq\rq}, J. Phys.Chem. {\bf 51}, 631 (1947).

\bibitem{Verwey-Overbeek-1948}  E.J.W. Verwey and J.T.G. Overbeek, {\it Theory of the Stability of Lyophobic Colloids}
 (Elsevier, Amsterdam, 1948).

\bibitem{Casimir-Polder-1946} H.B.G. Casimir and D. Polder, \lq\lq Influence of retardation on the London-van der Waals forces",
Nature 158, 787-788 (1946).

\bibitem{Casimir-Polder-1948} H.B.G. Casimir and D. Polder, \lq\lq The influence of retardation on the London-van der Waals forces'',
 Phys. Rev. {\bf 73}, 360-372 (1948).

\bibitem{Lennard-Jones-1932} J.E. Lennard-Jones, \lq\lq Processes of adsorption and diffusion on solid surfaces. {\rq\rq}, Trans.Faraday Soc. {\bf 28}, 333-359 (1932).

\bibitem{Tabor-Winterton-Nature-1968} D. Tabor and R.H.S. Winterton, \lq\lq Surface forces: direct measurements of normal and retarded van der Waals forces{\rq\rq}, Nature {\bf 219}, 1120-1121 (1968).

\bibitem{Tabor-Winterton-1968} D. Tabor and R.H.S. Winterton, \lq\lq The direct measurement of normal and retarded van der Waals forces {\rq\rq}, {\it Proc. Roy. Soc. Lond. A}{\bf 312}, 435-450 (1969).

\bibitem{Axilrod-Teller-1943} B.M. Axilrod and E. Teller, \lq\lq Interaction of the van der Waals Type Between Three Atoms",
 J. Chem. Phys. {\bf 11}, 299 (1943).

\bibitem{Langbein-1974} Dieter Langbein, {\it Theory of van der Waals Attraction},
 Springer Tracts in Modern Physics, vol. 72 (Springer-Verlag, Berlin, 1974).

\bibitem{Milonni-1994} Peter W. Milonni, {\it The Quantum Vacuum: an Introduction to Electrodynamics}, (Academic Press, San Diego, CA, 1994).

\bibitem{Farina-Santos-Tort-1999} C.Farina, F.C.Santos, and A.C.Tort, \lq\lq A simple way of understanding the non-additivity of van der Waals dispersion forces'', Am.J.Phys. {\bf 67}, 344-349 (1999).

\bibitem{Holstein-2001} B.R. Holstein, \lq\lq The van der Waals interaction'', Am. J. Phys. {\bf 69} (4), 441-449 (2001).

\bibitem{Power-1964} E.A. Power, {\it Introductory Quantum Electrodynamics}, (Longmans, London, 1964).

\bibitem{Craig-Thiru-1998} D.P. Craig and T. Thirunamachandran, {\it Molecular Quantum Electrodynamics}, (Dover, New York, 1998).

\bibitem{Salam-2010} Akbar Salam, {\it Molecular Quantum Electrodynamics: Long-Range Intermolecular Interactions},
 (John Wiley $\&$ Sons, New Jersey, 2010).

\bibitem{Taddei-Mendes-Farina-2010} M.M. Taddei, T.N.C. Mendes and C. Farina, \lq\lq An introduction to dispersive interactions{\rq\rq}, Eur. J. Phys. {\bf 31}, 89-99 (2010).

\bibitem{Raskin-Kusch-1969} D. Raskin and P. Kusch, ``Interaction between a Neutral Atomic or Molecular Beam and a Conducting Surface'', Phys.Rew. {\bf 179}, 712-721 (1969)

\bibitem{SukenikEtAl-1993} C.I. Sukenik, M.G. Boshier, D. Cho, V. Sandoghdar and E.A. Hinds, ``Measurement of the Casimir-Polder Force'', Phys.Rew.Lett. {\bf 70}, 560-563 (1993)

\bibitem{LandraginEtAl-1996} A. Landragin, J.-Y. Courtois, G. Labeyrie, N. Vansteenkiste, C.I. Westbrook and A. Aspect,
 ``Measurement of the van der Waals Force in an Atomic Mirror'', Phys.Rew.Lett. {\bf 77}, 1464-1467 (1996)

\bibitem{Shimizu-2001} Fujio Shimizu, \lq\lq Specular Reflection of Very Slow Metastable Neon Atoms from a Solid Surface{\rq\rq},
 Phys. Rev. Lett. {\bf 86}, 987-990 (2001).

\bibitem{Dalibard-2002} A. Aspect and J. Dalibard. \lq\lq Measurement of the atom-wall interaction: from London to Casimir-Polder'',
 Séminaire Poincaré {\bf 1}, 67-78 (2002).

\bibitem{Parsegian-Book} V. Adrian Parsegian, {\it Van der Waals Forces: A Handbook for Biologists, Chemists, Engineers, and Physicists},
 (Cambridge University Press, New Yourk, 2006).

\bibitem{Elizalde-Romeo-1991} E. Elizalde and A. Romeo, \lq\lq Essentials of the Casimir effect and its computation{\rq\rq},
Am. J. Phys. {\bf 59}, 711 (1991).

\bibitem{Farina-BJP-2006} C. Farina, \lq\lq The Casimir effect: some aspects{\rq\rq}, Braz. J. Phys. {\bf 36}, 1137-1149 (2006).

\bibitem{Proceedings-Leipzig-1998}  H.B.G. Casimir, \lq\lq Some remarks on the history
of the so called Casimir effect{\rq\rq}, in the {\it Proceedings of the
Fourth Workshop on Quantum Field Theory under the Influence of
External Conditions}, pg 3, Ed. M. Bordag, World Scientific (1999).

\bibitem{Milton-Book-2004} K.A. Milton, {\it The Casimir effect: physical manifestations of zero-point energy}, (World Scientific, 2011).

\bibitem{Mostepanenko-Book-2009} Michael Bordag, Galina Leonidovna Klimchitskaya, Umar Mohideen and Vladimir Mikhaylovich Mostepanenko,
{\it Advances in the Casimir Effect},  (Oxford University Press, Oxford, 2009).

\bibitem{AutumnEtAl-Gecko-2002} K. Autumn, M. Sitti, Y.A. Liang, A.M. Peattle, W.R. Hansen, S. Sponberg, T.W. Kenny,
R. Fearing, J.N. Israelachvili and R.J.Full, ``Evidence for van der Waals adhesion in gecko state'', Proc. Nat. Acad. Sci. {\bf 99},
12252-12256 (2002)

\bibitem{LeeEtAl-Nature-2007} H. Lee, B.P. Lee, P.B. Messersmith, ``A reversible wet/dry adhesive
inspired by mussels and geckos'', Nature {\bf 448}, 338-341 (2007).

\bibitem{Lamoreaux-PhysicsToday-2007} Steve K. Lamoreaux, \lq\lq Casimir forces: still surprising after 60 years{\rq\rq},
 Phys. Today {\bf 60}, 40-45 (2007).

\bibitem{Milton-AJP} Kimball A. Milton, \lq\lq Resource Letter VWCPF-1: Van der Waals and Casimir-Polder forces{\rq\rq},
 Am. J. Phys. {\bf 79}, 697-711 (2011).

\bibitem{Buhmann-Welsch-2007} S.Y. Buhmann and D.G Welsch, \lq\lq Dispersion forces in macroscopic quantum electrodynamics{\rq\rq},
 Prog. Quant. Elect. {\bf 31}, 51-130 (2007).

\bibitem{Rowlinson-Book-2002} J.S. Rowlinson, {\it Cohesion: A Scientific History of Intermolecular Forces}, (Cambridge University Press, Cambridge, 2002).

\bibitem{Eberlein-Zietal-2007} C. Eberlein and R. Zietal, \lq\lq Force on a neutral atom near
conducting microstructures.{\rq\rq}, {\it Phys.Rev.} {\bf A 75}, 032516 (2007).

\bibitem{Eberlein-Zietal-2011} C. Eberlein and R. Zietal, \lq\lq Casimir-Polder interaction between a polarizable
particle and a plate with a hole {\rq\rq} Phys. Rev. A {\bf 83}, 052514 (2011).

\bibitem{Contreras-Eberlein-2009} A.M. Contrera Reyes and C. Eberlein, Phys. Rev. A {\bf 80}, 032901 (2009).

\bibitem{ReinaldoEtAl-2011} Reinaldo de Melo e Souza, W.J.M. Kort-Kamp, C. Sigaud and C. Farina,
 \lq\lq Finite-size effects and nonadditivity in the van der Waals interaction{\rq\rq}, Phys. Rev. A{\bf 84}, 052513 (2011).

\bibitem{ReinaldoEtAll-Proceeding-2012} Reinaldo de Melo e Souza, W.J.M. Kort-Kamp, C. Sigaud and C. Farina,
 \lq\lq Sommerfeld's image method in the calculation of van der Waals forces{\rq\rq}, to appear in the Proceedings of QFEXT11, arXiv:1201.5701.

\bibitem{Byron1992} Frederick W. Byron, W. Fuller, {\it Mathematics of Quantum and Classical Physics}, (Dover,
New York, 1992).

\bibitem{Griffiths1999} David J. Griffiths, {\it Introduction to Electrodynamics}, (Prentice Hall, 3ed, New
Jersey, 1999 ). See particularly pages 124-125.

\bibitem{Davydov-Book-1976} A.S. Davydov, {\it Quantum Mechanics} (Pergamon, 2nd ed. 1976).

\bibitem{Marvin-Toigo-1982} A.M. Marvin, F. Toigo, ``van der Waals interaction between a point particle and
a metallic surface. I. Theory'', Phys.Rev.A {\bf 25} 782-802 (1982).

\bibitem{Jhe-Kim-1995A} W. Jhe, J.W. Kim, ``Atomic energy-level shifts near a dielectric microsphere'',
Phys.Rev.A {\bf51} 1150 (1995).

\bibitem{Jhe-Kim-1995B} W. Jhe, J.W. Kim, ``Casimir-Polder energy shift of an atom near a metallic sphere'',
Phys.Lett.A {\bf197} 192-196 (1995).

\bibitem{BuhmannEtAl-2004} S.Y. Buhmann, H.T. Dung, D.-G. Welsch, ``The van der Waals energy of atomic systems
near absorbing and dispersing bodies'' J.Opt.B: Quantum Semiclass.Opt. {\bf 6} S127-S135 (2004).

\bibitem{Sambale-Buhmann-Scheel-2010} Agnes Sambale, Stefan Yoshi Buhmann and Stefan Scheel,
 \lq\lq Casimir-Polder interaction between an atom and a small magnetodielectric sphere{\rq\rq},
 Phys. Rev. A {\bf 81}, 012509 (2010).

\bibitem{EllingsenEtAl-2011} Simen A. Ellingsen, Stefan Yoshi Buhmann and Stefan Scheel,
\lq\lq 	Temperature-independent Casimir-Polder forces in arbitrary geometries{\rq\rq},
  	Phys. Rev. A{\bf 84}, 060501 (2011).

\bibitem{EllingsenEtAl-2012} Simen A. Ellingsen, Stefan Yoshi Buhmann and Stefan Scheel,
\lq\lq 	Casimir-Polder energy-level shifts of an out-of-equilibrium particle near a microsphere{\rq\rq},
 	Phys. Rev. A{\bf 85}, 022503 (2012).

\bibitem{Buhmann-PrivateComunication} S. Buhmann, private comunication.

\bibitem{Santos-Tort-2004} F.C. Santos, A.C. Tort, \lq\lq The electrostatic field of a point charge and an electrical dipole in the presence of a conducting sphere{\rq\rq}, Eur. J. Phys. {\bf 25}, 859-868 (2004).

\bibitem{Arfken1985} This symmetry can be shown from Green's theorem, stated in page 58 of George Arfken,
{\it Mathematical Methods for Physicists}, (Academic Press, San Diego, 1985). To demonstrate that
$G(\mathbf{r},\mathbf{r}')$ is indeed symmetric under ${\bf r}\, \longleftrightarrow\, {\bf r}^{\,\prime}$ use in that theorem $\phi(\mathbf{y})=G(\mathbf{r},\mathbf{y})$ and
$\psi(\mathbf{x})=G(\mathbf{x},\mathbf{r}')$ along with the Green's function properties and boundary
conditions.

\bibitem{Sommerfeld-1896} A. Sommerfeld, \lq\lq Über verzweigte Potentiale im Raum {\rq\rq}, Proc.London Math.Soc. {\bf 29}, 395-429 (1897).

\bibitem{MendesEtAl-2008} T.N.C. Mendes, F.S.S. Rosa, A. Tenorio and C. Farina, \lq\lq Dispersion forces
between an atom a perfectly conducting wedge {\rq\rq},
 J. Phys. A{\bf 41}, 164029 (2008).

\bibitem{BrevikEtAl-1998} I. Brevik, M. Lygren, and V.N. Marachevsky, \lq\lq Casimir-Polder effect for a perfectly conducting wedge {\rq\rq},
 Ann. Phys. (N.Y.) {\bf 267}, 134-142 (1998).

\bibitem{Lifshitz-1956} E.M. Lifshitz \lq\lq The theory of molecular attractive forces between solids{\rq\rq},  Sov. Phys  \textbf{JETP 2}, 73-83 (1956).
\\
E.M. Lifshtz and L.P. Pitaevskii, {\it Landau and Lifshtz Course of
Theoretical Physics: Statistical Physics Part 2},
Butterworth-Heinemann (1980).

\bibitem{DLP-1961} I. Dzyaloshinskii, E.M. Lifshitz and
L.P. Pitaevskii, \lq\lq General theory of van der Waals' forces {\rq\rq}, Soviet Physics Uspekhi, {\bf 4}, 153-176 (1961).

 \bibitem{Sabisky-Anderson-1973} E.S. Sabisky and C.H. Anderson, \lq\lq Verification of the Lifshitz theory of the van der Waals potencial using liquid-Helium films {\rq\rq},
Phys. Rev. A {\bf 7}, 790-806 (1973).

\bibitem{LevinEtAl-2010} M. Levin, A.P. McCauley, A.W. Rodriguez, M.T.H Reid and S.G. Johnson,
  \lq\lq Casimir repulsion between metallic objects in vacuum{\rq\rq}
   Phys. Re Lett. {\bf 105}, 090403 (2010).

\bibitem{McCauleyETAl-2011} Alexander P. McCauley, Alejandro W. Rodriguez, M. T. Homer Reid and Steven G. Johnson,
  \lq\lq Casimir repulsion beyond the dipole regime {\rq\rq}, arXiv:1105.0404v1 (2011).

\bibitem{Rodriguez-Lopez-2011}  Pablo Rodriguez-Lopez,
\lq\lq Casimir repulsion between topological insulators in the diluted regime{\rq\rq},
 Phys. Rev. B {\bf 84}, 165409-165415 (2011).

\bibitem{BostromEtAl-2012} Mathias Boström, Bo E. Sernelius, Iver Brevik and Barry W. Ninham
 \lq\lq Retardation turns the van derWaals attraction into a Casimir repulsion as close as 3 nm{\rq\rq},
  Phys. Rev. A{\bf 85}, 010701 (2012).

\bibitem{Bostrom-Sernelius-2012} Mathias Boström and Bo E. Sernelius,
 \lq\lq Repulsive van der Waals forces due to hydrogen exposure on bilayer graphene{\rq\rq},
  Phys. Rev. A{\bf 85} 012508 (2012).

\bibitem{Shajesh-Shaden-2012} K.V. Shajesh and M. Schaden,
 \lq\lq Repulsive long-range forces between anisotropic atoms and dielectrics{\rq\rq},
 Phys. Rev. A{\bf 85}, 012523 (2012).

\bibitem{Feinberg-Sucher-1968} G. Feinberg and J. Sucher,
 \lq\lq General form of the retarded van der Waals potentials{\rq\rq},
  J.Chem. Phys. {\bf 48}, 3333-3334 (1968).

\bibitem{Feinberg-Sucher-1970} Gerald Feinberg and Joseph Sucher,  	
 \lq\lq General Theory of the van der Waals Interaction: A Model-Independent Approach{\rq\rq},
  Phys. Rev. A{\bf 2}, 2395 (1970).

\bibitem{Farina-Santos-Tort-2002AJP} C. Farina, F.C. Santos and A.C.Tort,
 \lq\lq A simple model for the nonretarded dispersive force between an electrically polarizable atom and
 a magnetic polarizable one{\rq\rq},
  Am.J.Phys. {\bf 70,} (2002)

\bibitem{Farina-Santos-Tort-2002} C. Farina, F.C. Santos, A.C. Tort,
 \lq\lq The non-retarded dispersive force between an electrically polarizable atom and a magnetically polarizable one{\rq\rq},
 J. Phys. A{\bf 35}, 2477 (2002).

\bibitem{SpagnoloEtAl-2007} S. Spagnolo, D.A.R. Dalvit and P.W. Milonni, \lq\lq Van der Waals Interactions in a Magneto-Dielectric Medium{\rq\rq},
Phys. Rev. A{\bf 75}, 052117 (2007).

\bibitem{Scheel-Buhmann-2009} Stefan Yoshi Buhmann and Stefan Scheel, \lq\lq Macroscopic quantum electrodynamics -
concepts and applications{\rq\rq}, Journal-ref: Acta Physica Slovaca {\bf 58}, 675 (2008).

\bibitem{Boyer-1969} T.H. Boyer,  \lq\lq Asymptotic Retarded van der Waals Forces Derived from Classical Electrodynamics
 with Classical Electromagnetic Zero-Point Radiation{\rq\rq},
 Phys. Rev. {\bf 180}, 19-24 (1969).

\bibitem{Boyer-1974} T.H. Boyer, \lq\lq Van der Waals forces and zero-point energy for dielectric and permeable materials{\rq\rq}, Phys. Rev. A{\bf 9}, 2078-2084 (1974).

\bibitem{Dorota-1993} Dorota Kupiszewska, \lq\lq Repulsive Casimir effect: A one-dimensional model of the force
between dielectric and permeable plates{\rq\rq},
 J. Mod. Opt. {\bf 40}, 517-523 (1993).

\bibitem{MessinaEtAl-LateralCP-2009} Riccardo Messina, Diego A. R. Dalvit, Paulo A. Maia Neto, Astrid Lambrecht and Serge Reynaud,
 \lq\lq Dispersive interactions between atoms and nonplanar surfaces{\rq\rq}, Phys. Rev. A{\bf 80}, 022119 (2009).







\end{thebibliography}
\end{document}